\documentclass[aps, preprint, groupedaddress, floatfix]{revtex4}

\usepackage{epsfig}
\usepackage{graphicx}
\usepackage{amsmath}
\usepackage{tabularx}
\usepackage{rotating}
\usepackage{color}
\usepackage{multirow}

\newcommand{\parallelsum}{\mathbin{\|}}

\begin{document}

\title{Charge transfer driven emergent phenomena in oxide
heterostructures}

\author{Hanghui~Chen$^{1,2}$ and Andrew Millis$^{3}$}

\affiliation{
 $^1$NYU-ECNU Institute of Physics, New York University Shanghai, 200062
 China\\
 $^2$Department of Physics, New York University, New York, NY, 10002, USA\\
 $^3$Department of Physics, Columbia University, New York, NY, 10027, USA\\}
\date{\today}

\begin{abstract}
  Complex oxides exhibit many intriguing phenomena, including
  metal-insulator transition, ferroelectricity/multiferroicity,
  colossal magnetoresistance and high transition temperature
  superconductivity. Advances in epitaxial thin film growth techniques
  enable us to combine different complex oxides with atomic precision
  and form an oxide heterostructure. Recent theoretical and
  experimental work has shown that charge transfer across oxide
  interfaces generally occurs and leads to a great diversity of
  emergent interfacial properties which are not exhibited by bulk
  constituents. In this report, we review mechanisms and physical
  consequence of charge transfer across interfaces in oxide
  heterostructures. Both theoretical proposals and experimental
  measurements of various oxide heterostructures are discussed and
  compared. We also review the theoretical methods that are used to
  calculate charge transfer across oxide interfaces and discuss the
  success and challenges in theory. Finally, we present a summary and
  perspectives for future research.
\end{abstract}

\maketitle

\section{Introduction}

Artificial oxide heterostructures can now be grown with atomic
precision~\cite{Hwang2012}.  At oxide interfaces, charge transfer is a
very general and robust phenomenon. With electrons moving from one
oxide to the other, new charge configurations can be induced at the
interface. These charge configurations can be substantially
different from those found in bulk versions of the constituent
materials.  As a consequence, at oxide interfaces new electronic,
magnetic and orbital states emerge.  A classical example of emergent
phenomena in oxide heterostructures is the LaAlO$_3$/SrTiO$_3$
interface~\cite{Ohtomo2004}, where a high mobility two-dimensional
electron gas exhibiting magnetism and superconductivity is discovered
at the interface~\cite{Thiel2006, Reyren2007, Brinkman2007,
  Caviglia2008, Li2011, Bert2011}, while both LaAlO$_3$ and SrTiO$_3$
are wide band gap insulators. During the past decade, designing new
oxide heterostructures and seeking new interfacial phenomena have been
a focus of condensed matter physics~\cite{Tokura2008, Zubko2011, Chakhalian2014}. The
exciting new discoveries pose a challenge for theory: can we reliably
describe and predict charge transfer phenomena in oxide
heterostructures, in particular when constituting oxides are strongly
correlated?

In this report, we discuss charge transfer effects at oxide
interfaces.  We first distinguish three important mechanisms of charge
transfer in oxide heterostructures: 1) polarity difference; 2)
occupancy difference and 3) electronegativity difference. For the
first two mechanisms, we briefly discuss representative examples since
excellent reviews are already available~\cite{Huijben2009, Chen2010a,
  Pentcheva2010, Hwang2012}. We focus on the last mechanism and
present a comprehensive review of various examples and different
emergent phenomena arising from interfacial charge transfer. Next we
briefly describe theoretical methods that are widely used in
literature to calculate charge transfer in oxide heterstructures and
discuss the theoretical challenges pertinent to descriptions of charge
transfer in strongly correlated materials.  Finally we present a
summary and our perspectives for the field of oxide
heterostructures. Space limitations and the rapid development of the
field mean that the review can not be comprehensive.
We apologize to those whose work is not included here.

\begin{figure}[t]
\includegraphics[angle=0,width=0.85\textwidth]{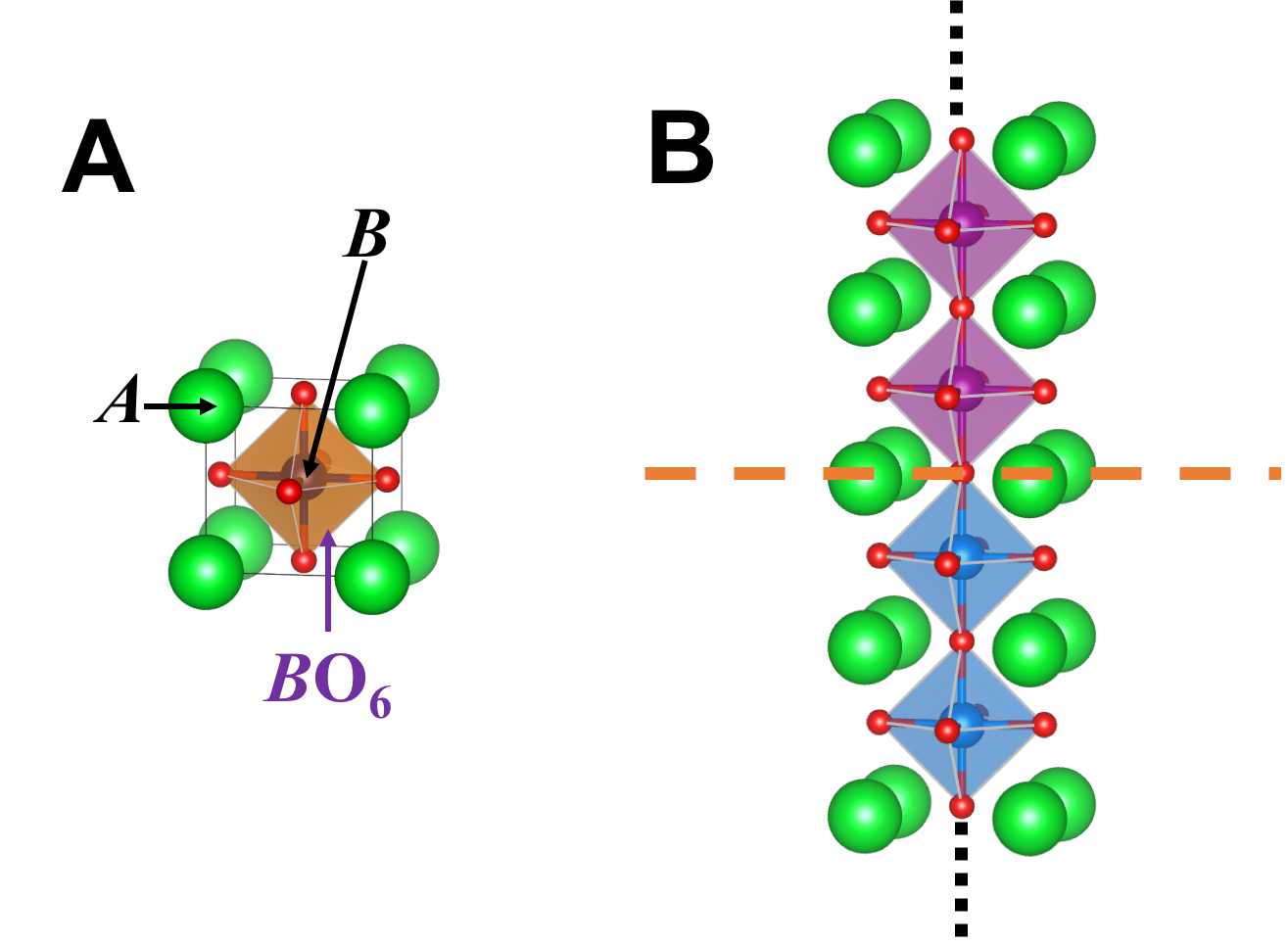}
\caption{\label{fig:perovskite} \textbf{A}) Atomic structure of a
  perovskite oxide (the formula unit is $AB$O$_3$). \textbf{B}) Two different transition metal oxides 
  are stacked along the [001] direction. The orange dashed line highlights the interface.}
\end{figure}

In this review we focus on an important class of transition
metal oxides: perovskite oxides. Their atomic structure is shown in
Fig.~\ref{fig:perovskite}\textbf{A}. The atom on the corner of the cube is called $A$-site atom,
which is either an alkaline earth metal or a rare earth metal. The atom at
the center of the cube is called $B$-site atom, which is a transition
metal. Each transition metal atom is surrounded by six oxygen atoms
which are at face-center of the cube. As we form an oxide
heterostructure using two perovskite oxides, we need to choose a
stacking direction. In this review, unless otherwise specified, 
we focus on (001) interfaces, which are shown in Fig.~\ref{fig:perovskite}\textbf{B}.

\section{Overview of charge-transfer mechanisms}

The materials separated by an interface will generically have
different electronic properties, and therefore different chemical potentials (measured, say, relative to the vacuum level) and this difference will generally
lead to charge flow across the interface. With the transferred electrons, the physical
and chemical properties of the constituent oxides close to the
interface can be fundamentally different from bulk properties because
the transition metal $d$ occupancy is changed.

In the context of oxide interfaces, it is useful to distinguish three
driving mechanisms all of which contribute to the chemical potential
difference: polarity difference, occupancy difference and
electronegativity difference. The classification of different charge
transfer mechanisms is not unambiguous. In fact, different mechanisms
are closely related and sometimes intertwined. The classification is
nonetheless useful, because charge transfer across oxide interfaces
always occurs to compensate for some type of ``discontinuity'', and
our classification lists the three most relevant types.

\subsection{Polarity difference}

In this review, a polar material is understood as an
insulator (polar metals have recently been experimentally
  synthesized~\cite{Shi2013, Kim2016a}, which however goes 
beyond the scope of our current discussion) such
that along a certain direction, in the form of a stoichiometric thin film, an average internal
electric field develops. Correspondingly, a nonpolar material is an
insulator such that in the stoichiometric thin film, along the given
direction, the internal electric field is averaged to zero. For
example, stoichiometric (001) LaAlO$_3$ films are polar because they
are composed of alternating (LaO)$^{+1}$ and (AlO$_2$)$^{-1}$ layers, while
stoichiometric (001) SrTiO$_3$ films are nonpolar since they consist of alternating
(SrO)$^0$ and (TiO$_2$)$^0$ layers. However, we note that along the
(110) direction, SrTiO$_3$ can be considered as polar because of the alternating
(SrTiO)$^{4+}$ and (O$_2$)$^{4-}$ layers.  

\begin{figure}[t]
\includegraphics[angle=0,width=0.85\textwidth]{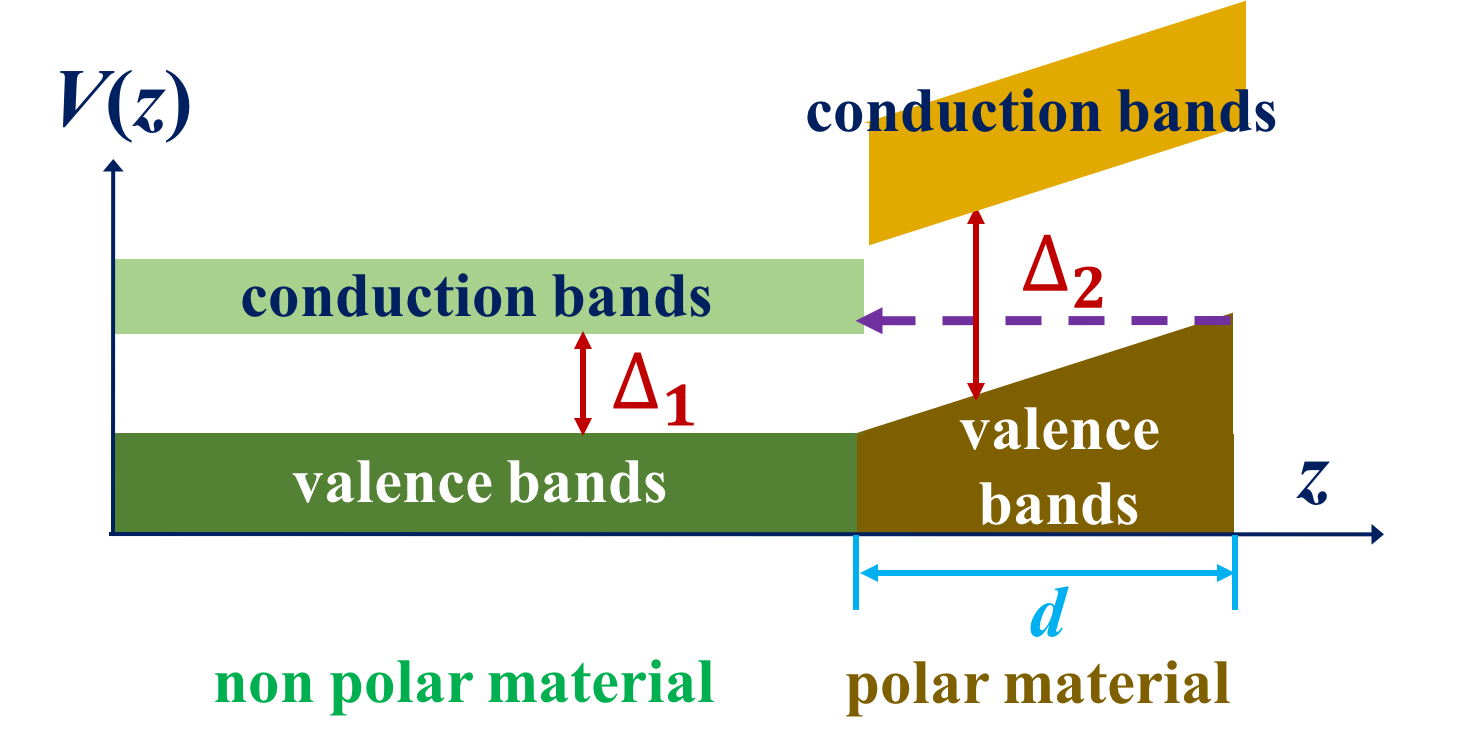}
\caption{\label{fig:polar} Potential profile of an ideal interface
  (i.e. no band misalignment) between a nonpolar material and a polar
  material. $\Delta_1$ ($\Delta_2$) is the band gap of the nonpolar
  (polar) material. $d$ is the thickness of the polar material.  In
  the polar material, an average internal polar field
  $\frac{dV}{dz} = E$ exists. The dashed arrow shows the charge
  transfer when the thickness $d$ is above the critical value (defined
  in Eq. (1) in the main text).}
\end{figure}

We focus on [001] as the stacking direction.
Fig.~\ref{fig:polar} illustrates the interface between a polar
material and a nonpolar material.  An average internal polar field
$E=\frac{dV}{dz}$ is developed in the polar material along the [001]
direction. The potential difference between one side of the polar
material and the other side is proportional to the thickness of the
material $d$. Therefore as

\begin{equation}
\label{eqn:eq1}  eEd > \rm{min}\{\Delta_1, \Delta_2\}
\end{equation}
electrons can tunnel from the surface to the interface ($\Delta_1$ is
the band gap of the nonpolar material and $\Delta_2$ is the band gap of
the polar material). As a
consequence of charge transfer, electrons emerge in the conduction band of
the nonpolar material (if $\Delta_1 < \Delta_2$) or in the conduction
band of the polar material (if $\Delta_1 > \Delta_2$) and holes appear in the valence band of the polar
material (on the surface). The smallest value of $d$ that satisfies Eq.~(\ref{eqn:eq1})
defines the critical thickness. For the $n$-type LaAlO$_3$/SrTiO$_3$
interface, the experimental critical thickness is 4 unit cells~\cite{Thiel2006}. As $d$
is above the critical thickness, the two-dimensional electron/hole gas
at the interface and surface counteracts the
internal field in the polar material. The sheet density of
electrons/holes increases with the thickness $d$ of the polar
material and approaches the saturation value as $1/d$ when the internal polar field is completely
compensated~\cite{Chen2010, Millis2010, Mannhart2008}. 

However, we need to make two important comments:

1) we note that Eq.~(\ref{eqn:eq1}) is based on the assumption that valence
bands are perfectly aligned. If there is a significant band misalignment
between the polar and the nonpolar materials, Eq.~(\ref{eqn:eq1})
needs to be refined. However, the general picture that charge transfers from
one side of the polar material to the other side above a critical
thickness remains qualitatively the same.

2) the polar catastrophe mechanism provides one way to compensate the
internal polar field. However, as the thickness $d$ of the polar
material is large enough (in the limit of bulk materials), other
compensation mechanism will be in play, such as vacancies, interstitials and
adsorbed molecules. The $\frac{1}{d}$ thickness dependence of sheet carrier density  
only applies to the situation without defect formation.

\subsection{Occupancy difference}

\begin{figure}[t]
\includegraphics[angle=0,width=0.85\textwidth]{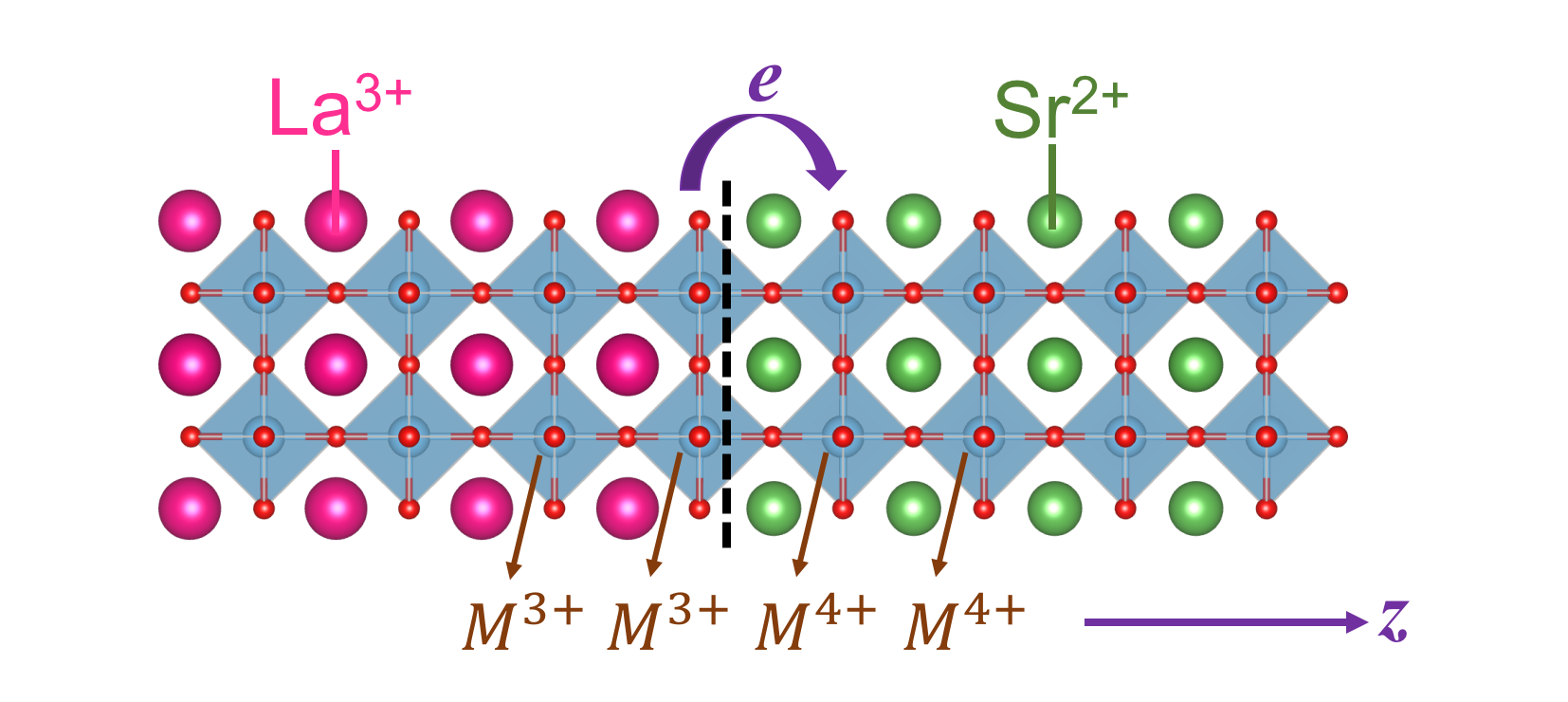}
\caption{\label{fig:occupancy} Atomic structure of the interface
  between La$M$O$_3$ and Sr$M$O$_3$ where $M$ is a transition metal
  ion. In La$M$O$_3$, the formal valence of $M$ is 3+, while in
  Sr$M$O$_3$, the formal valence of $M$ is 4+. The purple arrow
  indicates that electrons may transfer from $M^{3+}$ to $M^{4+}$ at
  the interface.}
\end{figure}

The second mechanism is the difference of transition metal $d$
occupancy across the interface. Many transition metal ions, such as
Ti, V and Mn, have multiple valences and the valence state can be
controlled by other elements in a compound.  For example, in $AB$O$_3$
perovskite materials, one may think of the O ions as having formal
valence $2^{-}$, while the $A$ ion may have formal valence $2^{+}$ (if
$A$=Sr, Ca, Ba, etc) or $3^{+}$ (if $A$=La or other member of the
lanthanide series), so charge neutrality fixes the valence of the
$B$-site transition metal ions as $4^{+}$ (if $A$ = Sr) or $3^{+}$ (if
$A$ = La). While formal valence is an oversimplification of the true
situation, it provides a useful description. Fig.~\ref{fig:occupancy}
illustrates the interface between La$M$O$_3$ and Sr$M$O$_3$. Without
charge transfer, the formal valence of transition metal ion $M$
abruptly changes from 3+ to 4+ at the interface.  This discontinuity
drives a charge flow, so that some electrons on $M^{3+}$ ions may flow
to $M^{4+}$ ions, which smoothes out the occupancy discontinuity at
the interface.  However, we note that difference in transition metal
$d$ occupancy does not always lead to charge transfer, or even if
charge transfer does occur, conduction does not necessarily emerge at
the interface. This is ascribed to the competition between correlation
effects and kinetic energies. We will discuss it in more details in
subsequent sections.

\subsection{Electronegativity difference}

\begin{figure}[t]
\includegraphics[angle=0,width=0.85\textwidth]{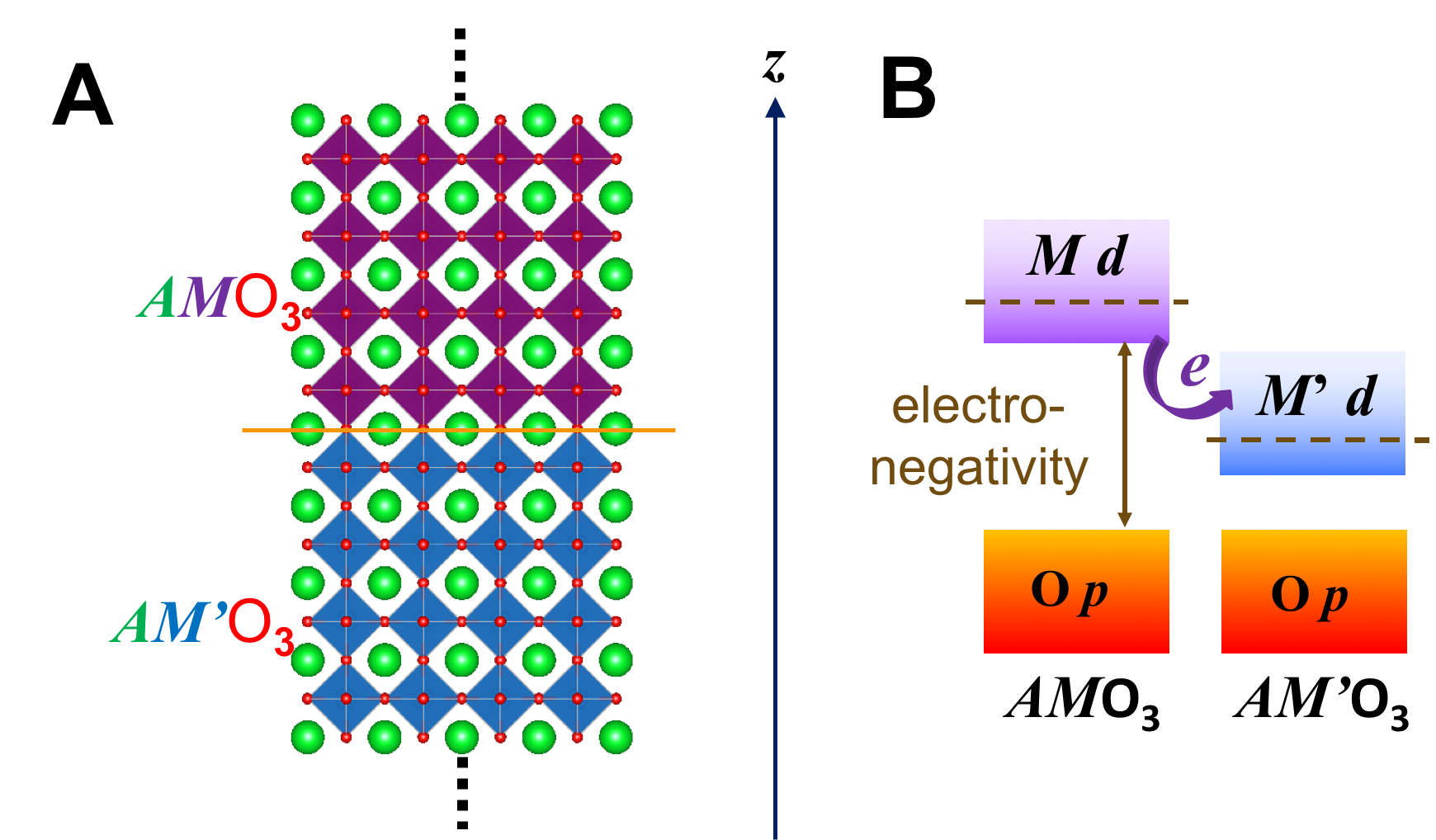}
\caption{\label{fig:electro} \textbf{A}) Atomic structure of the
  interface between $AM$O$_3$ and $AM'$O$_3$. \textbf{B}) Schematics
  of band alignments of $d$ states of transition metal $M$ and $M'$ as
  well as oxygen $p$ states. The solid arrow indicates the energy
  difference between metal $d$ states and oxygen $p$ states, a measure
  of electronegativity of transition metal $M$. If the $d$ states of
  transition metal $M$ have higher energy than those of $M'$,
  electrons can transfer from $M$ to $M'$ across the interface.}
\end{figure}

The third mechanism is the difference in electronegativity of
dissimilar transition metal $M$. Loosely speaking, electronegativity
(or sometimes referred to as electron affinity) is a measure of the
energy gain (or cost) of moving an electron from a reservoir to the
ion in question; of course the value of the electronegativity depends
on the choice of reservoir and on the valence of the ion. The
electronegativity, defined at constant valence, decreases as one moves
from left to right across the transition metal series and the
differences in electronegativity play an important role in the
magnitude of the charge transfer. As a rough rule of thumb, we observe
that the greater the electronegativity difference, the greater the magnitude of the charge
transfer.  To make these considerations more specific and
quantitative, we first remark that in transition metal oxides, one 
may think of the O$^{2-}$ as providing the reservoir. Thus the
electronegativity is in essence the energy separation between
transition metal $d$ and oxygen $p$ states, which is often referred to
as the charge transfer energy \cite{Zaanen1985}.  In many oxide
superlattices, the oxygen states approximately align across the
interface so that the electronegativity difference translates directly
into a contribution to the chemical potential difference and can drive
charge transfer, as shown in Fig.~\ref{fig:electro}\textbf{A}. This
figure shows the interface between $AM$O$_3$ and $AM^{\prime}$O$_3$,
where the two transition metal ions $M$ and $M^{\prime}$ have
identical formal valences (i.e. no polar discontinuity), but have
different energy levels of their $d$ states (different
electronegativity) as shown in Fig.~\ref{fig:electro}\textbf{B},
leading to transfer of electrons from $M$ to $M^{\prime}$ across the
interface to reduce the total energy. 

We note that similar to the previous discussion, we perfectly align the O
$p$ states across the oxide interface in Fig.~\ref{fig:electro}, which
is an oversimplification: although the oxygen states form a continuous
network the energies do not exactly align across interfaces.
However, the simplified picture provides a
very useful way of understanding the results of detailed calcuations.

\section{Polarity difference (Polar catastrophe)}

The charge transfer mechanism of polarity difference at oxide
interfaces is commonly known as the ``polar catastrophe''. The term
becomes commonly used after Ohtomo and Hwang synthesized LaAlO$_3$
thin films of a few unit cells thick on SrTiO$_3$ substrates with a
TiO$_2$ termination and discovered a high-mobility electron gas at the
LaAlO$_3$/SrTiO$_3$ interface~\cite{Ohtomo2004}. While the
LaAlO$_3$/SrTiO$_3$ interface (to be more precise, the one with
LaO/TiO$_2$ termination, sometimes referred to as the $n$-type
interface) is just one example of the ``polar catastrophe''
mechanism~\cite{Nakagawa2006}, it is an important special case which
has stimulated numerous theoretical and experimental works, and led to
many unexpected phenomena including magnetism~\cite{Brinkman2007,
  Li2011, Bert2011}, superconductivity~\cite{Reyren2007, Caviglia2008}
and tunable Rashba spin-orbital interaction~\cite{Caviglia2010,
  Lesne2016}. We refer readers to the excellent review papers that have already appeared in
literature~\cite{Huijben2009, Chen2010a, Pentcheva2010,
  Hwang2012}. 

However, we want to comment that while the LaAlO$_3$/SrTiO$_3$
interface is the prototype of the``polar catastrophe'' mechanism,
accumulating evidence shows that the ``polar catastrophe'' mechanism
alone can not explain all the observed experimental results. For
example, early experiments failed to observe the average internal
polar field in LaAlO$_3$~\cite{Segal2009}. Later experiments do report
an internal polar field, but the magnitude is only 80
meV/\AA~\cite{Singh-Bhalla2011}, smaller than the first-principles
calculations by at least a factor of 2~\cite{Chen2009, Chen2010}. This
implies that in addition to charge transfer, other mechanisms can also
screen the internal polar field, leading to smaller values than
theoretical predictions. Recently a polarity-induced defect mechanism
was proposed to account for both conduction and magnetism at the
LaAlO$_3$/SrTiO$_3$ interface~\cite{Yu2014}. Chambers
\textit{et. al.}~\cite{Chambers2011} show that at the (100)
LaCrO$_3$/SrTiO$_3$ interface, a potential gradient within the polar
material LaCrO$_3$ is sufficient to trigger a charge transfer, which
one would expect to lead to conduction. However, the interface is
experimentally found to be insulating. The insulating behavior was
attributed to cation-intermixture.

All these results show that at a general polar-nonpolar interface, the
``polar catastrophe'' picture which is based on the \textit{ideal}
atomic structure probably is not the only mechanism in play. Atomic
reconstruction, such as cation intermixture, various types of
vacancies and point defects, are very likely to occur. 

\section{Occupancy difference}

The charge transfer mechanism of occupancy difference between oxide
interfaces are closely related to the ``polar catastrophe''
mechanism. However, we use this classification to refer to one
particular type of superlattices which have been under intensive
study. The general formula of those superlattices can be expressed as
$RM$O$_3$/$AM$O$_3$ where $R$ is a tri-valent cation, $A$ is a
di-valent cation and $M$ is a transition metal ion. The most common
case for $R$ is La and for $A$ is Sr. Important examples of these
oxide heterostructures include LaTiO$_3$/SrTiO$_3$~\cite{Ohtomo2002,
  Okamoto2004}, LaMnO$_3$/SrMnO$_3$~\cite{Monkman2012, May2009} and
LaVO$_3$/SrVO$_3$~\cite{Luders2009, Dang2013} superlattices.

Like the LaAlO$_3$/SrTiO$_3$ interface, there are also many reviews in
literature discussing LaTiO$_3$/SrTiO$_3$, LaMnO$_3$/SrMnO$_3$ and
related oxide heterostructures~\cite{Pentcheva2010}. Here we briefly review these important
examples and mention some points that from our perspective deserve
attention for future research. 

Ref.~\cite{Ohtomo2002} shows that as a few unit cells of Mott
insulator LaTiO$_3$ are embedded into a band insulator SrTiO$_3$
matrix, electrons move from the Ti atoms in LaTiO$_3$ to the Ti atoms
in SrTiO$_3$, providing emergent conduction at the interface. Similar
phenomena have also been reported for a few unit cells of Mott
insulating GdTiO$_3$ embedded into an SrTiO$_3$ matrix.  The carrier
concentration at this interface is even higher~\cite{Moetakef2011}.
Ref.~\cite{Okamoto2004} shows that charge transfer and conduction are
general features at the interface between a semi-infinite Mott
insulator and a semi-infinite band insulator. However, if we change
the geometry, different phenomena can emerge. Refs.~\cite{Jang2011}
shows that if we insert only a single $R$O layer in a SrTiO$_3$
matrix, and if $R$ = La, Pr and Nd, conduction appears at the
interface, but if $R$ = Sm and Y, the interface remains
insulating. Ref.~\cite{Moetakef2012, Chen2013c} study another related
geometry: they consider inserting SrO in a Mott insulator GdTiO$_3$
matrix and both theory and experiment find that due to extreme quantum
confinement, a dimer Mott insulating state can be stabilized.

The second example is (LaMnO$_3$)$_m$/(SrMnO$_3$)$_n$ superlattices
with different Sr/La ratio (by varying $m$ and $n$). An important case
is that $m=2n$, which deserves special attention~\cite{Smadici2007}. For
(LaMnO$_3$)$_{2n}$/(SrMnO$_3$)$_n$ superlattices, as $n$ increases
from 1 to 5, a metal-insulator transition occurs (for $n\leq 2$, the
interface is metallic and for $n \geq 3$, the interface becomes
insulating)~\cite{Bhattacharya2008}. For the nature of insulating
states, Ref.~\cite{Bhattacharya2008} suggests that a finite peak does exist in the density of states at the
Fermi level but it is localized by disorder. A recent experiment~\cite{Monkman2012}
proposes that it is the quantum fluctuation that disrupts the
coherence of metallic states, giving rise to the insulating properties
observed in $n \geq 3$ superlattices. However, theoretical
work~\cite{Dong2008, Nanda2009} shows
that within a reasonable range of parameters, the ideal interface between
semi-infinite LaMnO$_3$ and semi-infinite SrMnO$_3$ should be metallic.
 
In Ref.~\cite{Luders2009} the authors synthesize
(LaVO$_3$)$_m$/(SrVO$_3$)$_1$ superlattices ($m$ varies from 2 to 6)
and find that the superlattices have a net magnetization up to room
temperatures due to the geometrically confined doping. However, 
the authors of Ref.~\cite{Dang2013, Dang2013a} show theoretically that the
experimentally determined crystal structure of LaVO$_3$/SrVO$_3$
superlattices is not favorable to induce ferromagnetism. They propose
that large amplitude of oxygen octahedral rotations would be needed to
stabilize a ferromagnetic state.

We note that for these (La$M$O$_3$)$_m$/(Sr$M$O$_3$)$_n$ superlattices
(where $M$ = Ti, Mn and V),
their chemical composition is equivalent to La$_{1-x}$Sr$_xM$O$_3$
where $x=\frac{n}{m+n}$ and $0<x<1$. The physical properties of solid
solution La$_{1-x}$Sr$_xM$O$_3$ have also been comprehensively
investigated in theory and experiment~\cite{Imada1998a, Salamon2001}. 
For $x=0$ or 1, the end material is usually insulating
(band insulator or Mott insulator). In the solid solution ($0 < x <
1$), conduction emerges within a range of $x$. However, as we have
seen from the above examples, with the same chemical composition, the
superlattices can exhibit distinct properties from solid
solutions. For examples, solid solution La$_{2/3}$Sr$_{1/3}$MnO$_3$ is ferromagnetic
metallic but (LaMnO$_3$)$_{10}$/(SrMnO$_3$)$_5$ superlattice is
insulating. (LaVO$_3$)$_6$/(SrVO$_3$)$_1$ superlattice exhibits a
large magnetic moment of 1.4 $\mu_B$/V ion, whereas solid solution 
La$_{6/7}$ Sr$_{1/7}$VO$_3$ shows a much smaller net magnetization. 
Another important difference of superlattices from compositionally
equivalent solid solutions is the anisotropy of transport. While solid solutions usually show
three dimensional conduction, the emergent conduction in superlattices
is confined to interfaces and exhibits two dimensional character,
which may be more useful for device development.

We also need to mention that in many systems, both charge occupancy difference and polarity effects
will contribute to the charge transfer. The LaAlO$_3$/SrTiO$_3$ and
LaTiO$_3$/SrTiO$_3$~\cite{Ohtomo2002}, GdTiO$_3$/SrTiO$_3$
~\cite{Moetakef2011a} systems provide useful examples. In the former,
the Al valence is the same in all layers and the charge transfer
depends strongly on the thickness of the polar LaAlO$_3$ layers,
becoming negligible for less than 4 unit cells, so here the charge
transfer is entirely driven by polarization effects. In the latter
systems the amount of charge transfer depends on the thickness of polar
LaTiO$_3$ (or GdTiO$_3$) layers but does not vanish even for 1
monolayer. Furthermore, the near interface Ti ions in the LaTiO$_3$
(or GdTiO$_3$) have a valence different from that of the Ti farther
from the interface; thus both mechanisms contribute in this system.

\section{Electronegativity difference}

\begin{figure}[t]
\includegraphics[angle=0,width=0.85\textwidth]{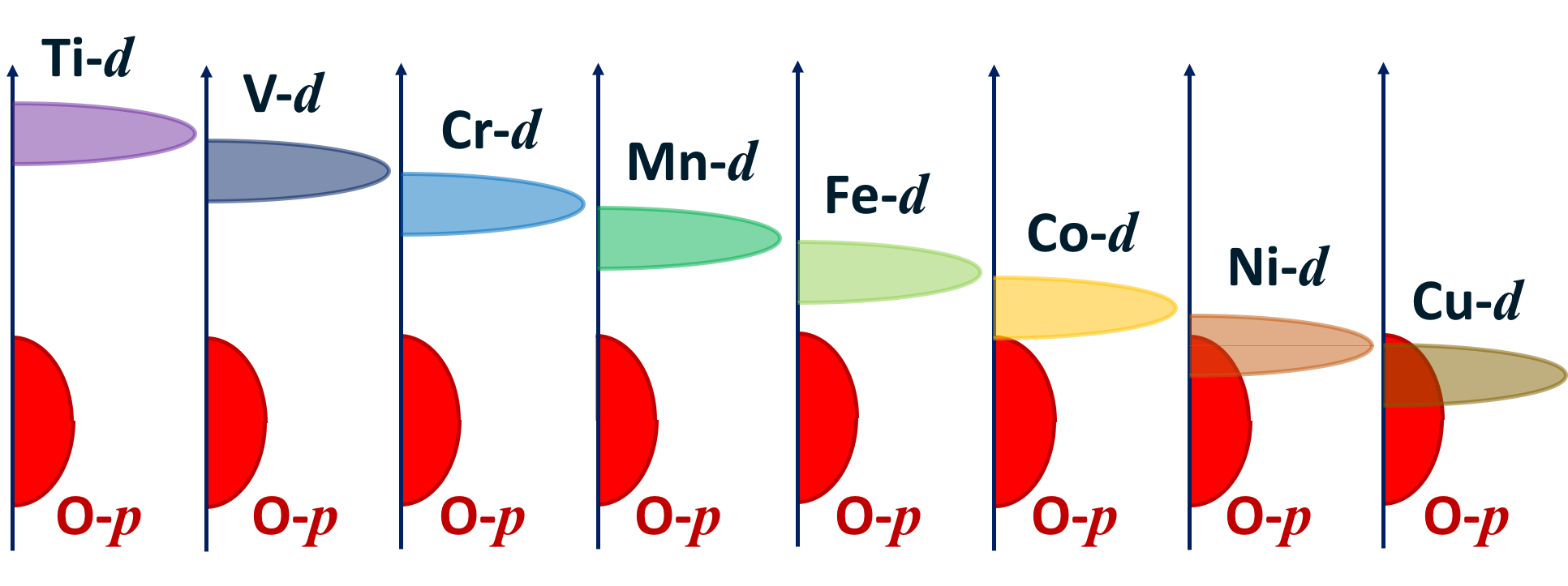}
\caption{\label{fig:d-states} Schematics of energy levels of
  transition metal $d$ states with respect to oxygen $p$ states in
  transition metal oxides  La$M$O$_3$ ($M$ = Ti, V, Cr, Mn, Fe, Co, Ni
  and Cu). As the mass of transition metal elements
  increases, the metal $d$ level decreases. For titanates, Ti-$d$
  states lie above O-$p$ by about 3 eV. For nickelates and cuprates, 
  Ni-$d$ and Cu-$d$ states even lie below O-$p$ states, leading to a
  ``negative charge transfer'' energy and strong hybridization.}
\end{figure}

In this section, we present a detailed review of
electronegativity-driven charge transfer in oxide
heterostructures. We focus on the following interfaces
$AM$O$_3$/$AM'$O$_3$ or double perovskite $A_2MM'$O$_6$, where $A$ is
di-valent or tri-valent ion and $M$, $M'$ are dissimilar transition
metal ions. Fig.~\ref{fig:d-states} schematically shows the energy
separation between metal $d$ and oxygen $p$ states for $3d$ transition
metal oxides La$M$O$_3$ ($M$ = Ti, V, Cr, Mn, Fe, Co, Ni and Cu). 
We note that as the mass of transition metal element
increases, the Coulomb attraction to the nucleus lowers the energy of
transition metal $d$ states, so the electronegativity increases as one
move from left to right along the transition metal row. For
titanates, the energy separation between Ti-$d$ and O-$p$ states is
about 3 eV. However, for nickelates and cuprates, the Ni-$d$ and
Cu-$d$ states lie even below the O-$p$ states, which leads to a ``negative
charge transfer'' energy and strong hybridization between metal $d$ and
oxygen $p$ states. The electronegativity-driven charge transfer is
based on the energy difference between the $d$ states of two different
transition metal ions.

Before we move on, we note that an electronegativity difference does
not always drive charge transfer. If one material $AM$O$_3$ is a wide
gap insulator with a nominally empty $M$-$d$ shell and the $d$ states
of $M$ ions lie above those of $M'$ ions, then no charge transfer
occurs between $M$ and $M'$ ions. SrTiO$_3$/SrVO$_3$ is one example in
which there is no charge transfer between SrTiO$_3$ and
SrVO$_3$~\cite{Yoshimatsu2010}. LaAlO$_3$/LaNiO$_3$ is another
example~\cite{Boris2011}. This situation is called quantum
confinement which reduces the band width of transition metal ions and
also leads to emergent phenomena, but it goes beyond the scope of our
current paper and we refer the readers to other review
papers~\cite{Stemmer2013}.

\subsection{LaTiO$_3$/LaNiO$_3$ and LaTiO$_3$/LaNiO$_3$/insulator superlattices}

\begin{figure}[t]
\includegraphics[angle=0,width=0.6\textwidth]{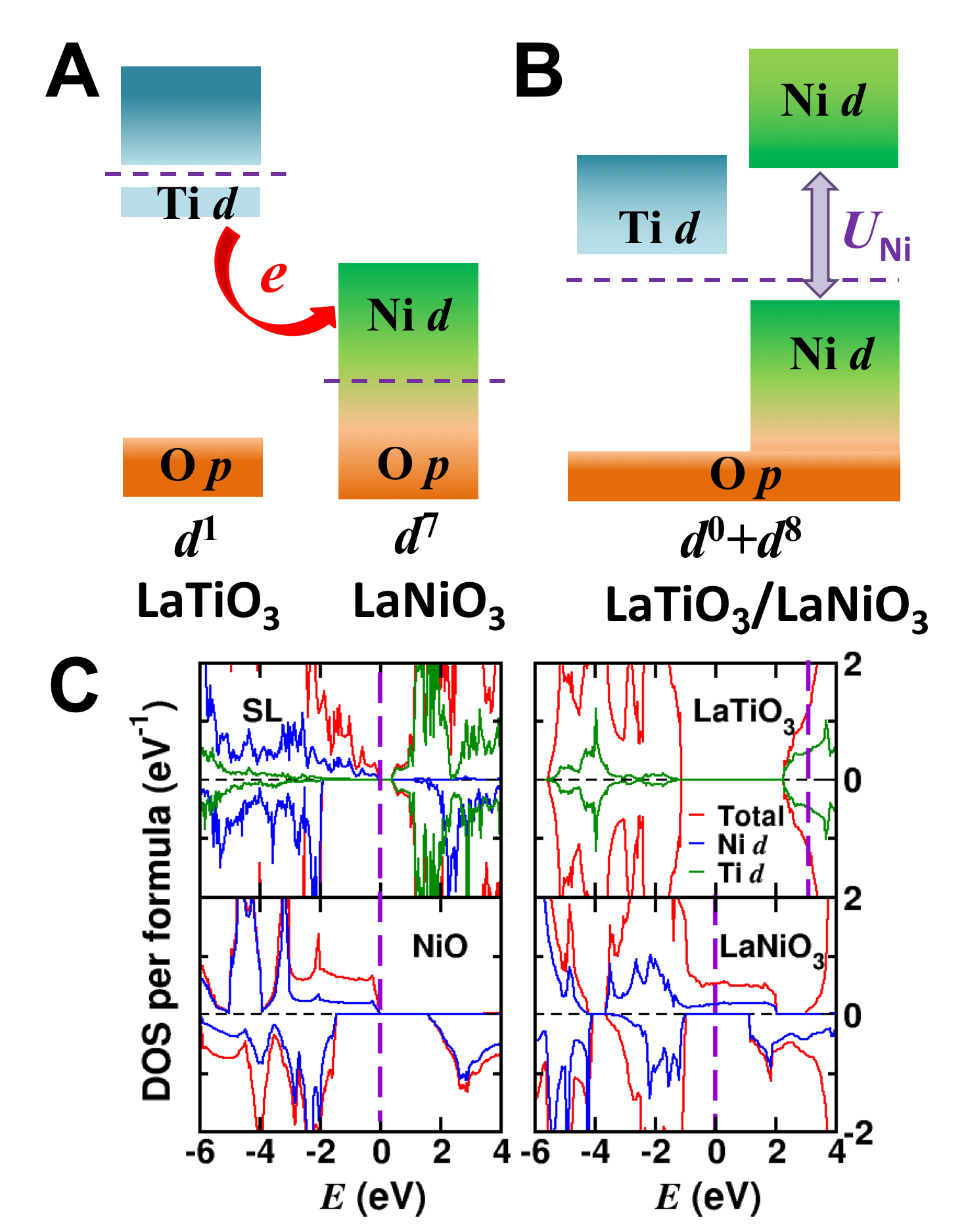}
\caption{\label{fig:ltolno} \textbf{A}) Schematic band structures of
  component materials LaTiO$_3$ and LaNiO$_3$. The dashed purple lines
  are the Fermi levels for the two materials. LaTiO$_3$ shows
  insulating behavior with a small excitation gap set by Ti $d$-$d$
  transitions and a wide energy separation between Ti $d$ states and O
  $p$ states.  LaNiO$_3$ exhibits metallic behavior with strong mixing
  between Ni $d$ states and O $p$ states. The red arrow highlights the
  direction of charge transfer in the superlattice. \textbf{B})
  Schematic band structure of (LaTiO$_3$)$_1$/(LaNiO$_3$)$_1$
  superlattice. Ti $d$ states are above the Fermi level (dashed purple
  line).  Correlation effects split Ni $d$ states into lower and upper
  Hubbard bands, separated by $U_{\textrm{Ni}}$. \textbf{C}) Densities
  of states for majority (above axis) and minority (below axis) spins
  of superlattice (upper left) and reference materials NiO (lower
  left), LaTiO$_3$ (upper right; zero of energy is shifted so that
  oxygen bands align with those of LaNiO$_3$) and LaNiO$_3$ (lower
  right). The densities of states are obtained using DFT+$U$
  calculations with $U_{\textrm{Ni}}$ = 6 eV and $U_{\textrm{Ti}}$ = 4
  eV. This figure is taken from Ref.~\cite{Chen2013a}.}
\end{figure}

We start from the interface between LaTiO$_3$ and LaNiO$_3$. Bulk
LaTiO$_3$ is an antiferromagnetic $S=1$ Mott insulator with a nominal
$d^1$ occupancy on Ti atoms. Bulk LaNiO$_3$ is a paramagnetic metal
with a nominal $d^7$ occupancy on Ni atoms. As
Fig.~\ref{fig:ltolno}\textbf{A} shows, in bulk LaTiO$_3$, Ti-$d$
states lie above O-$p$ states by about 3 eV, while in bulk LaNiO$_3$,
Ni-$d$ states have strong hybridization with O-$p$ states. In a
LaTiO$_3$/LaNiO$_3$ superlattice with a short periodicity, the lone electron on Ti-$d$
states is expected to transfer to Ni-$d$ states. Therefore after the
charge transfer, in the superlattice Ti atoms nominally have a $d^0$
occupancy and Ni atoms nominally have a $d^8$ occupancy. In the new
charge configuration, as Fig.~\ref{fig:ltolno}\textbf{B} shows, Ti
atoms have an empty $d$ shell. Ni atoms have a full $t_{2g}$ shell and
a half-filled $e_g$ shell. If the correlation strength on Ni sites is strong enough,
a Mott gap will open up. In this way, by design, we
can induce an artificial Mott insulating state via a nominally
complete charge transfer from Ti to Ni in LaTiO$_3$/LaNiO$_3$
superlattices~\cite{Chen2013a}. Fig.~\ref{fig:ltolno}\textbf{C} presents the
theoretically calculated density of states for (LaTiO$_3$)$_1$/(LaNiO$_3$)$_1$
superlattices, as compared to the density of states for the classical Mott insulator NiO (in which Ni
atoms also nominally have a $d^8$ occupancy) as well as bulk LaNiO$_3$
and LaTiO$_3$. We can see that the Ti-$d$ conduction bands, which are
partially filled in bulk LaTiO$_3$, become completely empty in the
(LaTiO$_3$)$_1$/(LaNiO$_3$)$_1$ superlattice. On the other hand, the
Ni-$d$ states, which are partially filled in bulk LaNiO$_3$, are
filled up in the (LaTiO$_3$)$_1$/(LaNiO$_3$)$_1$ superlattice. As a
consequence, a gap is opened in the superlattice, which separates
Ni-$d$ and Ti-$d$ states. The lower and upper Hubbard bands of Ni-$d$
states in the superlattice are very similar to those in NiO, which is
a strong evidence of the charge-transfer-driven Mott insulating state
on Ni sites.

\begin{figure}[t]
\includegraphics[angle=0,width=\textwidth]{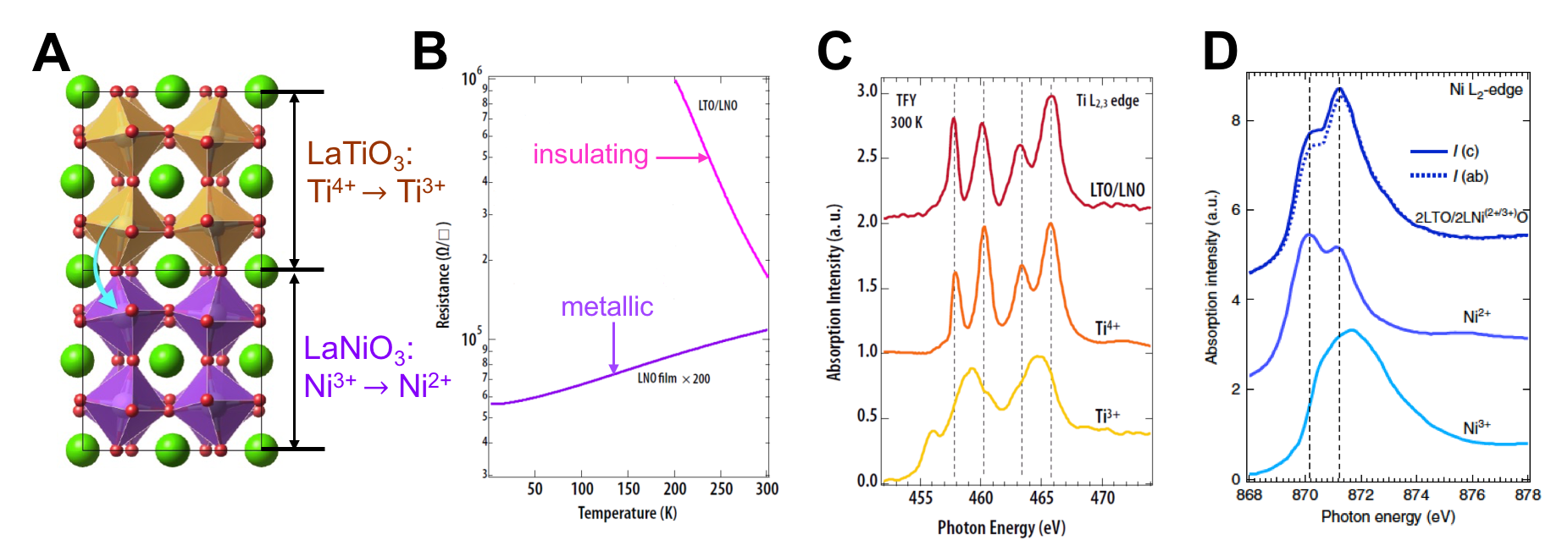}
\caption{\label{fig:ltolno-exp} \textbf{A}) Atomic structure of
  (LaTiO$_3$)$_2$/(LaNiO$_3$)$_2$ superlattices. \textbf{B}) 
Temperature-dependent sheet resistances of the
(LaTiO$_3$)$_2$/(LaNiO$_3$)$_2$ superlattices and the reference LaNiO$_3$ film
(20 unit cells). It is noteworthy that the sheet resistance of
LaNiO$_3$ film is $\times 200$. \textbf{C}) and \textbf{D}) X-ray
absorption spectroscopy (XAS) of
  (LaTiO$_3$)$_2$/(LaNiO$_3$)$_2$ superlattices. \textbf{C}) Ti
  $L_{2,3}$-edge. The reference spectra for Ti$^{4+}$ and Ti$^{3+}$
  were measured on a SrTi$^{4+}$O$_3$ single crystal and
  YTi$^{3+}$O$_3$ film ($\sim$ 100 nm on TbScO$_3$ substrate),
  respectively. \textbf{D}) Ni $L_{2,3}$-edge. The reference samples
  are bulk Ni$^{2+}$O and LaNi$^{3+}$O$_3$. Out-of-plane ($I$(c), dark
  blue solid line, $E$ $\parallelsum$ c and $E$ is the linear polarization vector of
  the photon) and in-plane ($I$(ab), dark blue dashed line, $E$ $\parallelsum$ ab)
  linearly polarized x-ray were used to measure XAS of (LaTiO$_3$)$_2$/(LaNiO$_3$)$_2$ superlattices at Ni
  $L_{2,3}$-edge. Black dashed lines are guidelines for peak positions. All
  spectra were collected and repeated more than two times with
  bulk-sensitive total fluorescence yield (TFY) detection mode at room
  temperature. This figure is adapted from Ref.~\cite{Cao2016}.}
\end{figure}

Inspired by the theoretical predictions in Ref.~\cite{Chen2013a}, Cao
et al.~\cite{Cao2016} synthesized a (LaTiO$_3$)$_2$/(LaNiO$_3$)$_2$
superlattice and measured the transport
properties. Fig.~\ref{fig:ltolno-exp}\textbf{A} shows the atomic
structure of the superlattice. Fig.~\ref{fig:ltolno-exp}\textbf{B}
shows the temperature dependence of sheet resistance of
(LaTiO$_3$)$_2$/(LaNiO$_3$)$_2$ superlattice and LaNiO$_3$ film.  The
former is highly insulating while the latter exhibits metallic
behavior. To probe the change in charge states,
Fig.~\ref{fig:ltolno-exp}\textbf{C} and \textbf{D} show the Ti
$L_{2,3}$ and Ni $L_2$ edge of (LaTiO$_3$)$_2$/(LaNiO$_3$)$_2$
superlattices measured by the x-ray absorption spectroscopy.  Panel
\textbf{C} shows that the formal valence of Ti in the superlattice
changes from the value 3+ in bulk LaTiO$_3$ to 4+ as in SrTiO$_3$,
while the formal valence of Ni in the superlattice changes from the
value 3+ in bulk LaNiO$_3$ to 2+ as in NiO. These results are
consistent with the theoretical predictions in Ref.~\cite{Chen2013a}
and provide convincing evidence of charge transfer in the
(LaTiO$_3$)$_2$/(LaNiO$_3$)$_2$ superlattices.

In addition to the charge transfer from Ti to Ni across the interface
between titanates and nickelates, Grisolia et. al.~\cite{Grisolia2016}
find a more subtle effect. By synthesizing and comparing
GdTiO$_3$/$R$NiO$_3$ interfaces ($R$ = La, Nd, Sm), they find that the
magnitude of charge transfer from Ti to Ni can be tuned by the element
$R$. The charge transfer increases from LaNiO$_3$ to NdNiO$_3$ to
SmNiO$_3$. The underlying mechanism is that different ionic sizes of
$R$ affect the oxygen octahedral rotations and thus hybridization
between Ni-$d$ and O-$p$ states. Because what we for simplicity refer
to as the $d$-state is an antibonding $p$-$d$ hybrid, the $p$-$d$
covalency affects the energy. Therefore, in addition to the energy
gain by moving electrons from Ti-$d$ states of higher energies to
Ni-$d$ states of lower energies, Ni-$d$ and O-$p$ states change their
hybridization and covalent character (so-called ``rehybridization''),
which also costs energy. The larger the ionic size, the more energy the
rehybridization costs, which limits the amount of
charge that can be transferred across the interface. It is
noteworthy~\cite{Grisolia2016} that while hybridization in $R$NiO$_3$
tunes the charge transfer from Ti to Ni across the interface between
GdTiO$_3$ and $R$NiO$_3$, hybridization in $R$TiO$_3$ plays a much
less significant effect on the charge transfer across the interface
between $R$TiO$_3$ and LaNiO$_3$. The reason is that different from
strong hybridization between Ni-$d$ and O-$p$ states, in titanates
Ti-$d$ states lie above O-$p$ states by about 3 eV, which results in a
much weaker hybridization between Ti-$d$ and O-$p$ states. Therefore,
changing the ionic size of $R$ in $R$TiO$_3$ can not significantly
tune the hybridization and therefore does not have the controlling
effects on charge transfer as eminent as it does in $R$NiO$_3$.
    
\begin{figure}[t]
\includegraphics[angle=0,width=\textwidth]{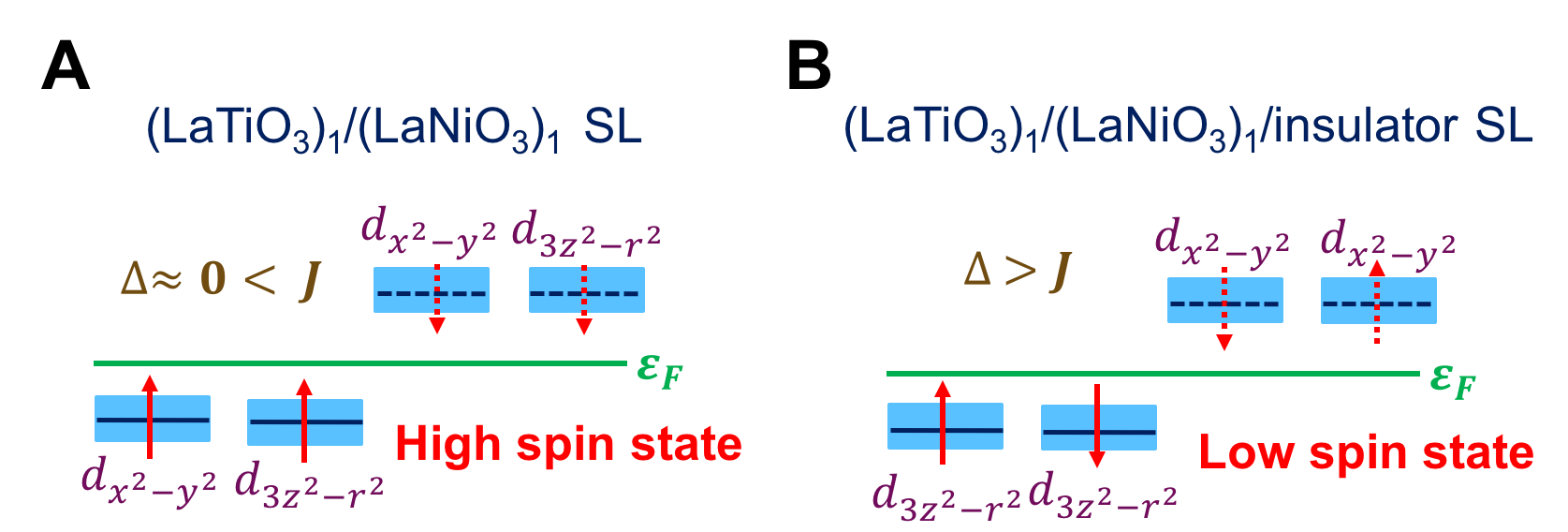}
\caption{\label{fig:high-low} Charge and spin configurations of
  \textbf{A}) (LaTiO$_3$)$_1$/(LaNiO$_3$)$_1$ superlattice and
  \textbf{B}) (LaTiO$_3$)$_1$/(LaNiO$_3$)$_1$/insulator
  superlattice. $J$ is the Hund's coupling for Ni $d$ states and
  $\Delta$ is the orbital splitting between Ni $d_{x^2-y^2}$ and
  $d_{3z^2-r^2}$ orbitals. $\varepsilon_F$ is the Fermi level. The shallow
  blue patches illustrate band widths and the dark blue solid/dashed
  line highlight the central positions of bands. As
  $\Delta < J$, the two electrons fill two different orbitals with the same
  spin, leading to a $S=1$ high-spin state. As $\Delta > J$, the two
  electrons fill the same orbital with opposite
  spins, leading to a $S=0$ low-spin state.}
\end{figure}

Next we discuss a LaTiO$_3$/LaNiO$_3$/insulator tri-component
superlattice~\cite{Chen2013}. The motivation of designing such a new
superlattice is to engineer an unprecedented orbital state in Ni atoms
(in addition to the change in charge states)~\cite{Chaloupka2008a,
  Hansmann2009}. As in the LaTiO$_3$/LaNiO$_3$ superlattice,
nominally one electron transfers from Ti to Ni in the tri-color superlattice, which
leads to a Ni $d^8$ occupancy with a full $t_{2g}$ shell and two
electrons in the $e_g$ shell. However, they have fundamental
difference. As Fig.~\ref{fig:high-low} shows, in the LaTiO$_3$/LaNiO$_3$
superlattice, the two electrons in the Ni $e_g$ shell form a high-spin
$S=1$ state, while in the tri-component superlattice, the two electron
in the Ni $e_g$ shell form a low-spin $S=0$ state. The
high-spin/low-spin configuration is determined by the competition
between Hund's coupling $J$ and crystal field splitting $\Delta$, which is the energy
difference between Ni $d_{x^2-y^2}$ and $d_{3z^2-r^2}$ orbitals. If
$\Delta < J$, the two electrons fill two different orbitals with the
same spin, leading to a high-spin state. If $\Delta > J$, the two
electrons fill the same orbital with opposite spins, leading to a
low-spin state. The significance of a low-spin state in Ni-based oxide
heterostructures is that further electron doping of the tri-component
superlattice can induce a single-orbital Fermi surface, analogous to
that of superconducting cuprates. 

\begin{figure}[t]
\includegraphics[angle=0,width=\textwidth]{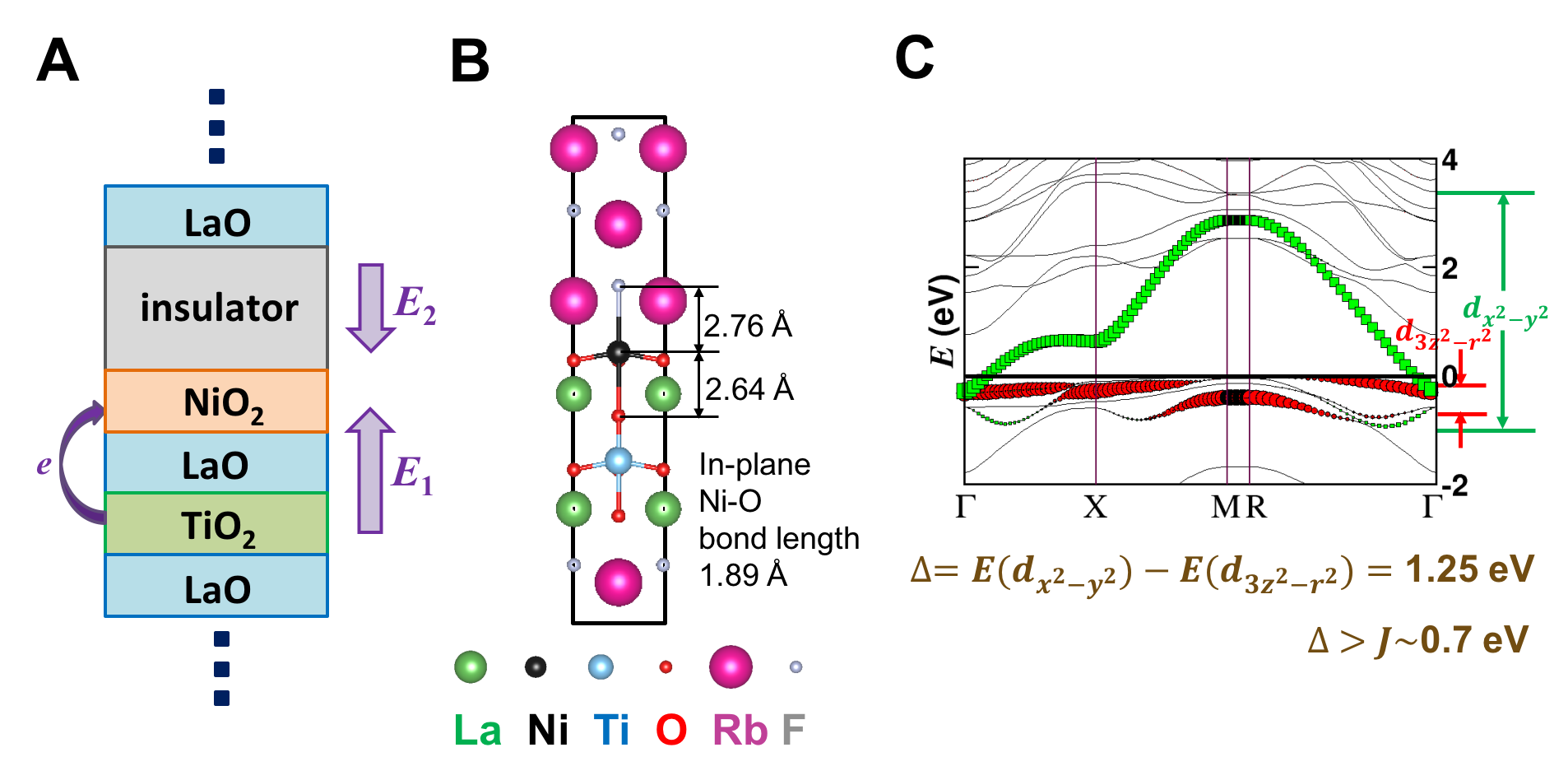}
\caption{\label{fig:tri-color} \textbf{A}) Schematics of
  LaNiO$_3$/LaTiO$_3$/insulator superlattices. \textbf{B})
  Theoretically calculated atomic structure of (LaNiO$_3$)$_1$/(LaTiO$_3$)$_1$/(RbF)$_3$
  superlattices. The in-plane Ni-O bond length is 1.89 \AA. The
  out-of-plane Ni-O and Ni-F bond lengths are 2.64 and 2.76 \AA. \textbf{C}) Band structure of
  (LaNiO$_3$)$_1$/(LaTiO$_3$)$_1$/(RbF)$_3$ superlattice. The red
  symbols are band projections onto Ni $d_{3z^2-r^2}$ orbital.  The green
  symbols are band projections onto Ni $d_{x^2-y^2}$ orbital. Using
  the Wannier functions to fit the DFT-calculated band structure, the
  difference between the on-site energy for Ni $d_{3z^2-r^2}$ orbital
  and the on-site energy for Ni $d_{x^2-y^2}$ orbital is found to be 1.25 eV. The
  Hund's coupling for Ni $d$ orbital is about 0.7 eV.}
\end{figure}

The orbital splitting $\Delta$ between $d_{x^2-y^2}$ and
$d_{3z^2-r^2}$ orbitals is induced by a Jahn-Teller-like distortion,
i.e. elongation of out-of-plane Ni-O bonds. Such a structural
distortion occurs to the tri-component superlattice by the
combination of charge transfer and insertion of a wide-gap
insulator. Fig.~\ref{fig:tri-color}\textbf{A} shows the schematics of
a (LaTiO$_3$)$_1$/(LaNiO$_3$)$_1$/insulator tri-component
superlattice. Charge transfer from Ti to Ni induces an internal
electric field $E_1$ and due to the periodic boundary condition a
second electric field $E_2$ appears. Both $E_1$ and $E_2$ pull the
apical oxygen atoms away from the Ni atom, which leads to Jahn-Teller-like
distortions that favor the occupancy of Ni $d_{x^2-y^2}$ orbital over
the $d_{3z^2-r^2}$ orbital. The presence of a wide gap insulator
explicitly breaks the inversion-symmetry and leads to $E_1\neq E_2$.
This asymmetry induces a polar distortion on Ni atoms, i.e. Ni and O
are not co-planar, which further increases the out-of-plane Ni-O bond
length. Fig.~\ref{fig:tri-color}\textbf{B} shows the theoretically calculated
atomic structure of (LaTiO$_3$)$_1$/(LaNiO$_3$)$_1$/(RbF)$_3$
superlattices. We note that the out-of-plane Ni-O and Ni-F bond
lengths are on average equal to 2.7~\AA, which is much longer than the
in-plane Ni-O bond length 1.89~\AA. The ferroelectric-like Ni-O
displacement is clearly visible. Fig.~\ref{fig:tri-color}\textbf{C}
shows the band structure of (LaTiO$_3$)$_1$/(LaNiO$_3$)$_1$/(RbF)$_3$
superlattice. The red symbols are band projections onto Ni
$d_{3z^2-r^2}$ orbital. The green symbols are band projections onto Ni
$d_{x^2-y^2}$ orbital. The Ni $d_{x^2-y^2}$ band is almost completely
filled up while the Ni $d_{3z^2-r^2}$ band is nearly empty. This
confirms that the tri-component superlattice indeed has a low-spin
configuration (two electrons fill the same orbital with opposite
spins). On the other hand, using the maximally localized Wannier
functions and fitting them to the DFT-calculated band structure, the
difference $\Delta$ between the onsite energy for Ni
$d_{3z^2-r^2}$ orbital and that for Ni $d_{x^2-y^2}$ orbital is
calculated to be 1.25 eV. The Hund's coupling $J$ for Ni $d$ orbital
is about 0.7 eV. Therefore it is $\Delta > J$, which is consistent
with the low-spin configuration.

\begin{figure}[t]
\includegraphics[angle=0,width=0.85\textwidth]{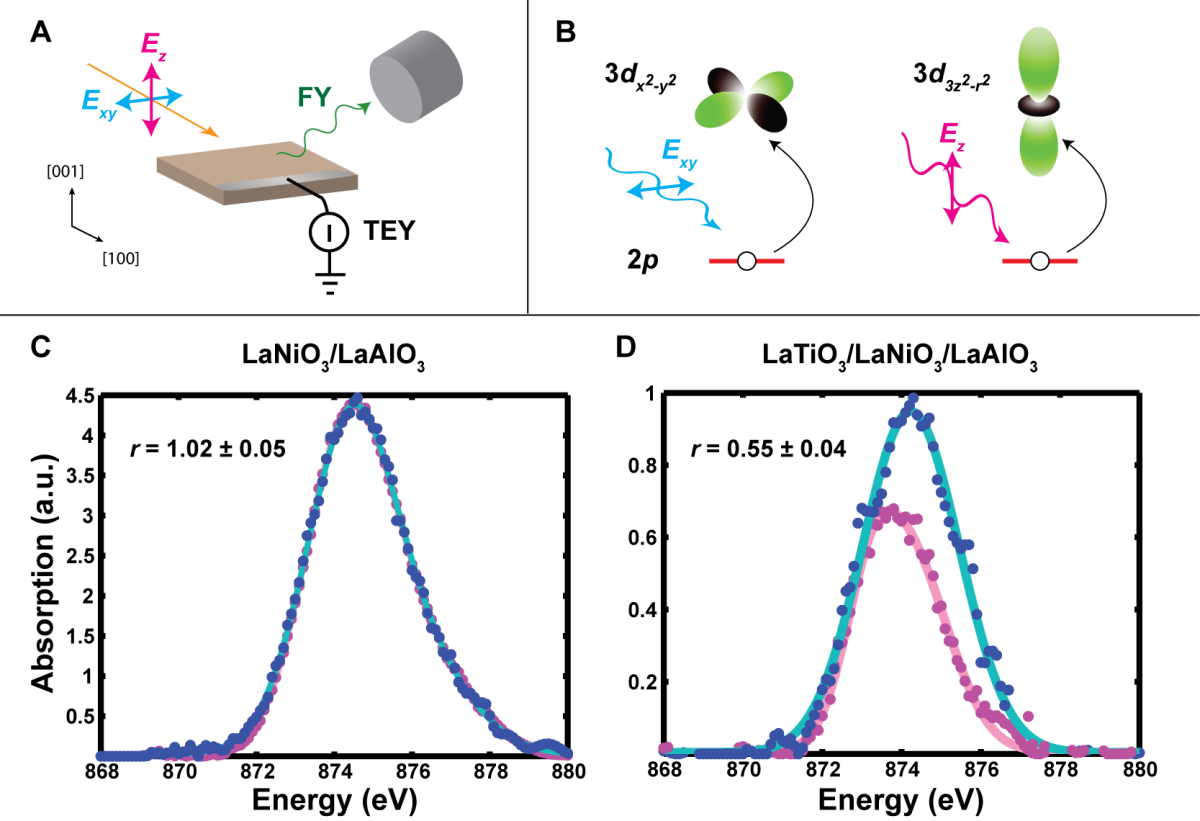}
\caption{\label{fig:tri-color-exp} Large orbital polarization in
  three-component superlattices observed by x-ray linear
  dichroism. \textbf{A}) Schematic of the experiment. \textbf{B}) The orbital
  selective atomic transitions probed by x-ray linear dichroism. \textbf{C})
  Measured x-ray absorption (circles) for in-plane (blue) and
  out-of-plane (pink) polarizations for a two-component and \textbf{D})
  three-component nickelate superlattice. The solid colored lines are
  double Gaussian fits. This figure is adapted from Ref.~\cite{Disa2015}.}
\end{figure}

Following the theoretical proposal of Ref.~\cite{Chen2013}, Disa et
al.~\cite{Disa2015} synthesized an artificial
(LaTiO$_3$)$_1$/(LaNiO$_3$)$_1$/(LaAlO$_3$)$_3$ superlattice and used
x-ray linear dichroism (XLD) to probe the orbital
occupancy. Fig.~\ref{fig:tri-color-exp}\textbf{A} shows the
experimental setup and Fig. ~\ref{fig:tri-color-exp}\textbf{B} shows
the orbital selective atomic transitions probed by the x-rays. The
experiments compared LaNiO$_3$/LaAlO$_3$ superlattice (two-component
superlattice) and the tri-component superlattice. The resulting
orbital-polarization-dependent spectra are shown in
Fig.~\ref{fig:tri-color-exp}\textbf{C} and \textbf{D}. In the
two-component superlattice, no significant dichroic signal is observed
(panel \textbf{C}). The red (absorption for Ni $d_{3z^2−r^2}$ orbital) and blue
(absorption for Ni $d_{x^2−y^2}$ orbital) symbols almost overlap with each
other. In contrast, there is a marked dichroism for the tri-component
superlattice (panel \textbf{D}). This result represents the largest
experimentally observed Ni $e_g$ orbital polarization in perovskite
nickelate systems to date. Furthermore, the Ni $e_g$ orbital occupancy
measured experimentally is in good agreement with first-principles DFT
calculations. However, G. Fabbris et al.~\cite{Fabbris2016} measured resonant
inelastic x-ray scattering (RIXS) spectra for this superlattice
recently and obtained good fits to the spectra by using a $d$-only
model, i.e. we consider $dd$ excitations from a Ni $d^8$ atom without explicitly including
oxygen $p$ states. The fits give an on-site energy splitting between the two Ni
$e_g$ orbital of only about 0.2 eV in the superlattice, whereas a $d$-only
Wannier function analysis of the DFT-calculated band structure gives
an on-site splitting of about 0.8 eV~\cite{Disa2015}. This discrepancy, along
with the fact that the $d$-only RIXS does yield the predicted orbital
polarization, has been attributed to the hybridization between Ni $d$
and O $p$ states ~\cite{Fabbris2016}. We believe that the $d$-only Wannier functions
which are used to fit the DFT band structure treat $p$-$d$ hybridization in
a different manner from the $d$-only RIXS model. Nevertheless, both the
DFT calculations and the experimental spectra (XLD and RIXS) find a
large orbital polarization in the (LaTiO$_3$)$_1$/(LaNiO$_3$)$_1$/(LaAlO$_3$)$_3$
superlattice. Further work is needed to address these remaining
issues.

\subsection{LaTiO$_3$/LaFeO$_3$ and YTiO$_3$/YFeO$_3$ superlattices}

LaTiO$_3$/LaFeO$_3$ and YTiO$_3$/YFeO$_3$ superlattices are another
example of a nominally complete charge transfer from Ti to Fe. The two
superlattices share similarities but also have important differences. 

\begin{figure}[t]
\includegraphics[angle=0,width=\textwidth]{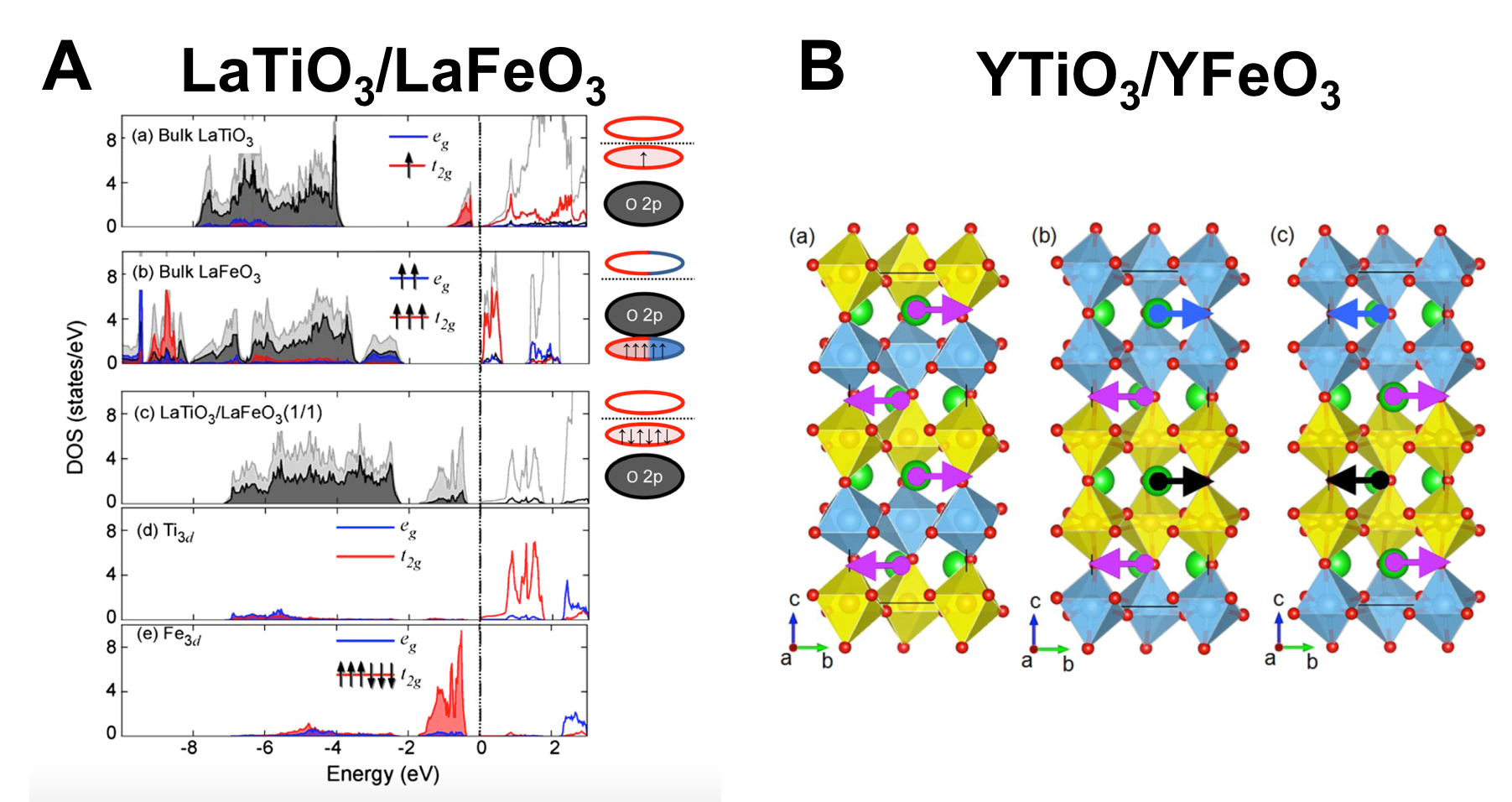}
\caption{\label{fig:Ti-Fe} \textbf{A}) Atomic and orbital projected
  density of states as well as schematic band structure of (a) bulk
  LaTiO$_3$, (b) bulk LaFeO$_3$, and (c,d,e)
  (LaTiO$_3$)$_1$/(LaFeO$_3$)$_1$ superlattice. Total states are
  marked in gray, O $p$ states in black, Fe and Ti $t_{2g}$ states in
  red, and Fe and Ti $e_g$ states in blue. The Fermi level is
  indicated by the dotted line. \textbf{B}) Atomic of
  (YTiO$_3$)$_n$/(YFeO$_3$)$_n$ superlattice ($n$ = 1 and 2). Sketch of ferroelectric
  distortions. The arrows denote the displacements of Y$^{3+}$ . (a) $n$ =
  1. The displacements are compensated between layers. The (b)
  positive and (c) negative ferroelectric distortion for $n$ = 2. This
figure is adapted from Ref.~\cite{Kleibeuker2014} and Ref.~\cite{Zhang2015}.}
\end{figure}

Bulk LaFeO$_3$ (YFeO$_3$) nominally has a Fe $d^5$ occupancy. The
half-filled Fe $d$ shell forms a high-spin state. Bulk LaTiO$_3$
(YTiO$_3$) nominally has a Ti $d^1$
occupancy. Ref.~\cite{Kleibeuker2014} shows both in theory and in
experiment that Ti atoms nominally donate one electron to Fe atoms
across the LaTiO$_3$/LaFeO$_3$ interface and after the charge transfer
Fe atoms nominally have a $d^6$ occupancy but a high-spin to low-spin
transition occurs and the six electrons completely fill the Fe
$t_{2g}$ shell. Fig.~\ref{fig:Ti-Fe}\textbf{A} illustrates such a
charge-transfer-driven spin transition. The theoretically calculated
density of states shows an empty Ti $t_{2g}$ shell and a fully
occupied Fe $t_{2g}$ shell. Experimentally, by using x-ray
photoelectron spectroscopy, the authors of Ref.~\cite{Kleibeuker2014}
confirmed the rearrangement of the Fe $3d$ bands and revealed an
unprecedented charge transfer up to $1.2 \pm 0.2 e^-$ per interface
unit cell in the LaTiO$_3$/LaFeO$_3$ heterostructures. In
YTiO$_3$/YFeO$_3$ superlattices, a similar charge transfer from Ti to
Fe is also found in theory~\cite{Zhang2015}. However, in contrast to
the LaTiO$_3$/LaFeO$_3$ superlattices, a robust high-spin state is
found in the YTiO$_3$/YFeO$_3$ superlattices, probably due to the
small ionic size of Y which leads to a smaller bandwidth of Fe $d$
states and favors the high-spin configuration. In addition to the
high-spin state, hybrid ferroelectricity with a polarization
$P\sim 1 \mu$C/cm$^2$ is induced in a (YTiO$_3$)$_2$/(YFeO$_3$)$_2$
superlattice. Fig.~\ref{fig:Ti-Fe}\textbf{B} shows the atomic
structure of (YTiO$_3$)$_n$/(YFeO$_3$)$_n$ superlattices ($n$=1 and 2)
with arrow highlighting the displacement of Y atoms. If $n=1$, all the
Y atoms have the same environment. If $n=2$, there are three types of
Y atoms: one is sandwiched between TiO$_6$ and FeO$_6$, one is
sandwiched between two TiO$_6$ and one is sandwiched between two
FeO$_6$. The displacements of all three types of Y atoms do not
exactly cancel each other, which leads to a net polarization.

We note that in Ref.~\cite{Kleibeuker2014, Zhong2016}, the authors align the
valence band edge of O-$2p$ and find the occupied Ti $t_{2g}$ states
have overlap with the empty upper Hubbard bands of Fe-$d$ states in
the energy window. However, in Ref.~\cite{Zhang2015}, the authors align
the localized O-$2s$ states and find the occupied Ti-$t_{2g}$ states
lie between the occupied lower Hubbard bands and the unoccupied upper
Hubbard bands of Fe-$d$ states. However, in the calculations of both
superlattices, charge transfer occurs from Ti to Fe atoms. This shows
that the rigid band alignment using the bulk band structures can only
serve as an approximate guide. Charge transfer
and the resulting band alignment at oxide interfaces should be
determined in a self-consistent way, as performed in superlattice calculations. 

\subsection{LaMnO$_3$/LaNiO$_3$ superlattices}

Next we discuss the LaMnO$_3$/LaNiO$_3$ superlattice. This is an
interesting case, since up till now there is inconsistency between
theory and experiment. While both theory and experiment indicate that 
charge transfer from Mn to Ni occurs due to electronegativity differences,
experimental transport and optical measurements show that
(LaMnO$_3$)$_2$/(LaNiO$_3$)$_2$ superlattices are
insulating~\cite{Hoffman2013, DiPietro2015} but
theoretical calculations find that the (LaMnO$_3$)$_m$/(LaNiO$_3$)$_n$
superlattices with different Mn/Ni ratios $m/n$ are all metallic~\cite{Lee2013}. 

\begin{figure}[t]
\includegraphics[angle=0,width=0.85\textwidth]{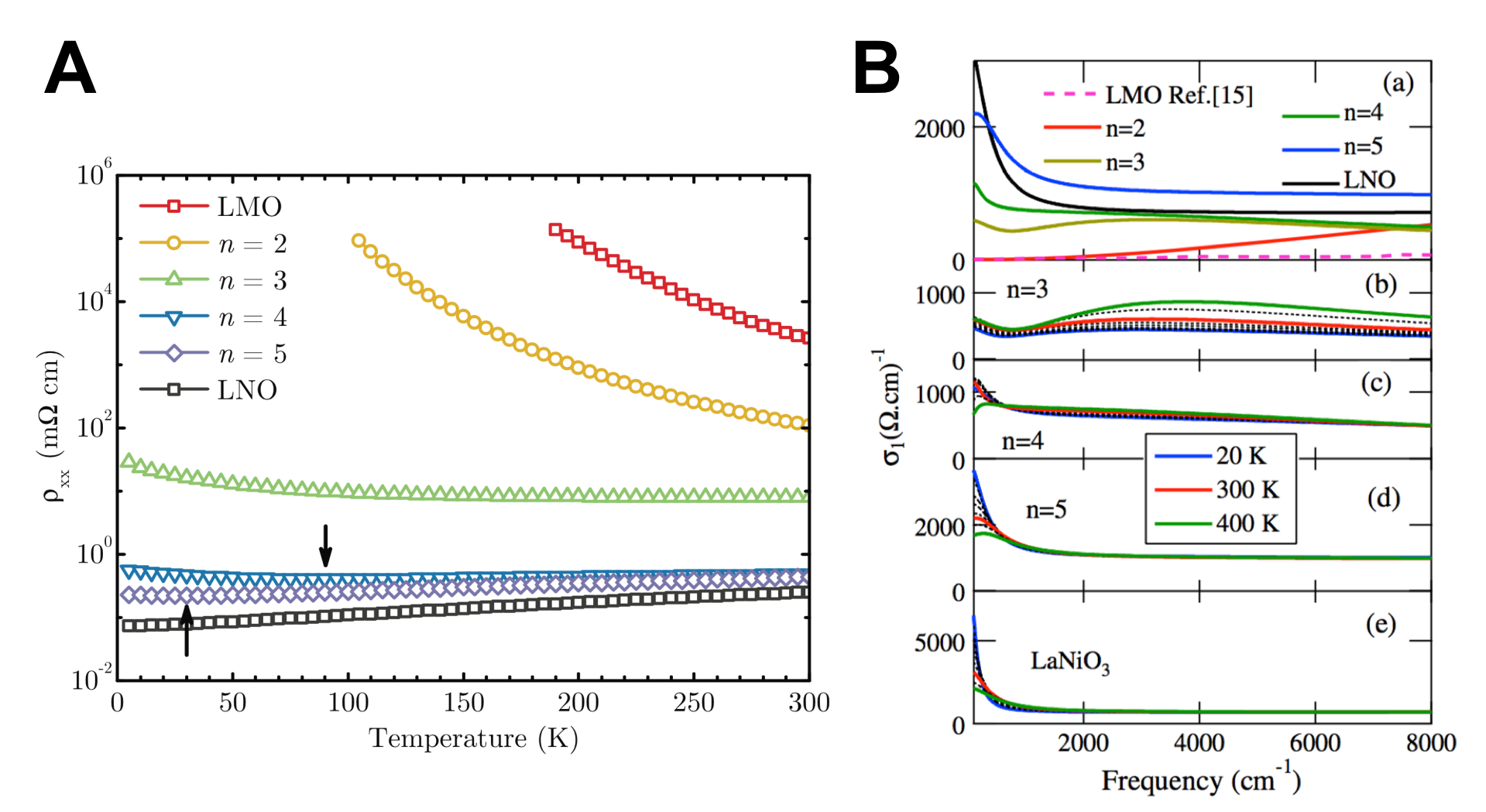}
\caption{\label{fig:Mn-Ni} \textbf{A}): Transport properties of 
(LaNiO$_3$)$_n$/(LaMnO$_3$)$_2$ superlattices ($2\leq n \leq 5$). 
Temperature dependence of longitudinal resistivity $\rho_{xx}$. The
arrows indicate positions of the resistivity minima at $T$ = 90K ($n =
4$) and $T$ = 30K ($n = 5$). \textbf{B}) Optical conductivity of the LaNiO$_3$/LaMnO$_3$
superlattices as extracted from a Lorentz-Drude fitting. (a) Room
temperature optical conductivities of the 
$n = 2,3,4,5$ samples and pure LaNiO$_3$ and LaMnO$_3$. (b)
Temperature dependence of the optical conductivity of the $n=3$
compound. (c,d,e) Same for the $n=4,5$ and for pure LaNiO$_3$. This
figure is adapted from Ref.~\cite{Hoffman2013} and Ref.~\cite{DiPietro2015}.}
\end{figure}

Fig.~\ref{fig:Mn-Ni}\textbf{A} shows the transport properties of
(LaMnO$_3$)$_2$/(LaNiO$_3$)$_n$ superlattices. As $n$ decreases from 5
to 2, a metal-insulator transition occurs. In particular, in
experiment the (LaMnO$_3$)$_2$/(LaNiO$_3$)$_2$ superlattice exhibits
strong insulating behavior. Fig.~\ref{fig:Mn-Ni}\textbf{B} shows the
optical conductivity of (LaMnO$_3$)$_2$/(LaNiO$_3$)$_n$ as a function
of $n$, temperature and frequency. Similar to transport measurements,
the low frequency optical conductivity substantially drops as $n$
decreases from 5 to 2. In both Ref.~\cite{Hoffman2013} and
Ref.~\cite{DiPietro2015}, the authors ascribe the observed
metal-insulator transition to the charge transfer from Mn to Ni. Such
a charge transfer in confirmed in first-principles
calculations~\cite{Lee2013}. However, for both
(LaMnO$_3$)$_1$/(LaNiO$_3$)$_1$ and (LaMnO$_3$)$_2$/(LaNiO$_3$)$_2$,
no insulating state is stabilized in the calculations. Tuning the
Hubbard $U$ for Mn and Ni $d$ orbitals in a reasonable range does not
change the metallic properties of the superlattice. This raises the
question whether the charge transfer from Mn to Ni is a nominally
complete charge transfer or not. If it is, then presumably a Mott
insulating state should emerge in (LaMnO$_3$)$_1$/(LaNiO$_3$)$_1$
superlattice where Mn atoms have a half-filled $t_{2g}$ shell and Ni
atoms have a full $t_{2g}$ and half-filled $e_g$ shell, similar to
(LaTiO$_3$)$_1$/(LaNiO$_3$)$_1$ superlattice. However, theoretical
calculations show that a partial charge transfer from Mn to Ni occurs
and the superlattice remains metallic. We note that double perovskite
La$_2$MnNiO$_6$ is found to be a ferromagnetic insulator in both
theory~\cite{Das2008} and experiment~\cite{Singh2007, Singh2009}.  The
inconsistency between theory and experiment on LaMnO$_3$/LaNiO$_3$
superlattices implies that the interface may not be atomically sharp
and disorder such as antisite defects~\cite{Chen2016} could play a
role in inducing the insulating state.  Further research, in
particular characterization of interfacial atomic structure using
high-resolution electron microscopy, may help to resolve
the problem.

We note here that for LaMnO$_3$/LaNiO$_3$ interfaces, in addition to
(001) stacking direction, (111) interfaces have also been synthesized
and studied~\cite{Gibert2012, Gibert2016}. Ref.~\cite{Gibert2012} shows that at (111)
LaMnO$_3$/LaNiO$_3$ interface, in addition to charge transfer,
exchange bias emerges, which implies the development of
interface-induced magnetism in the paramagnetic LaNiO$_3$ layers.
Such a bias does not show up at the (001) LaMnO$_3$/LaNiO$_3$ interface.

\subsection{Ba$_2$VFeO$_6$, Pb$_2$VFeO$_6$ and Sr$_2$VFeO$_6$ double perovskite oxides}

\begin{figure}[t]
\includegraphics[angle=0,width=\textwidth]{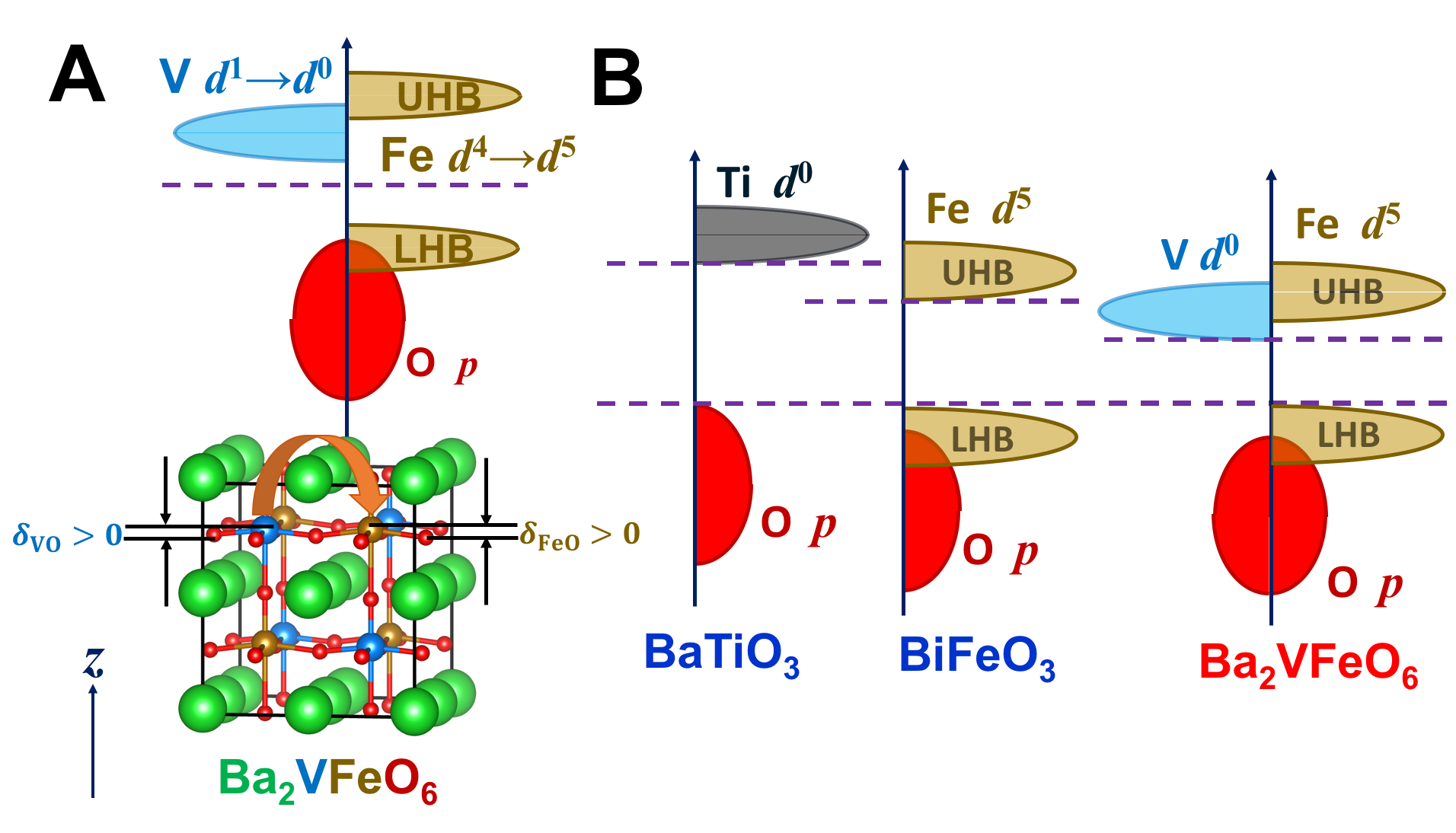}
\caption{\label{fig:V-Fe} \textbf{A}) Energy diagram and atomic
  structure of double perovskite Ba$_2$VFeO$_6$. The dashed line is the
  Fermi level, which lies in the gap between V $d$ and Fe $d$
  states. ‘LHB’ (‘UHB’) means lower Hubbard bands (upper Hubbard
  bands). The red arrow indicates the charge transfer from V atoms to
  Fe atoms due to electronegativity difference. In the double
  perovskite Ba$_2$VFeO$_6$, a polar distortion is developed
  ($\delta_{\textrm{VO}} > 0$  and $\delta_{\textrm{FeO}} > 0$) because of the
  new charge configuration V $d^0$ and Fe $d^5$. \textbf{B}) 
  Comparison of gaps for perovskite oxides: BaTiO$_3$, BiFeO$_3$ and
  Ba$_2$VFeO$_6$. The valence band edges are aligned for comparison.}
\end{figure}

Substantial charge transfer not only occurs to atomically sharp
interfaces in superlattices, but also in double perovskite which are
bulk compounds that are based on two single perovskite oxides (see
Fig.~\ref{fig:V-Fe} for the atomic structure). Double
perovskite Ba$_2$VFeO$_6$ is one example in which a nominally complete
transfer from V to Fe leads to Mott multiferroic properties which do
not exhibit in either bulk BaVO$_3$ or bulk BaFeO$_3$. Perovskite
BaVO$_3$ crystallizes in cubic structure with a nominally V $d^1$
occupancy. Perovskite BaFeO$_3$ also crystallizes in cubic structure
with a nominally Fe $d^4$ occupancy. As Fig.~\ref{fig:V-Fe}\textbf{A}
shows, in the double perovskite Ba$_2$VFeO$_6$, a nominally complete
charge transfer leads to new charge configurations V $d^0$ and Fe
$d^5$. Since Fe atoms have a half-filled configuration, strong
correlation is expected to open a Mott gap. At sufficiently low
temperatures, the large local magnetic moment $S=5/2$ on Fe atoms is
expected to order magnetically. More importantly,
both the empty V $d$ shell and the half-filled Fe $d$ shell have a ferroelectric
instability, just like the Ti $d^0$ state in BaTiO$_3$ and the Fe $d^5$ state
in BiFeO$_3$.  The presence of Ba ions which have a large ionic size
creates favorable conditions for ferroelectricity~\cite{Wu2011}. 
Fig.~\ref{fig:V-Fe}\textbf{B} compares the band structure of
BaTiO$_3$, BiFeO$_3$ and Ba$_2$VFeO$_6$. In BaTiO$_3$ and
Ba$_2$VFeO$_6$, both Ti and V have $d^0$ occupancy. However, due to the
electronegativity difference between Ti and V, the Ti $d$ states lie
above the V $d$ states, which leads to a smaller band gap for Ba$_2$VFeO$_6$
than for BaTiO$_3$. On the other hand, in both BiFeO$_3$ and
Ba$_2$VFeO$_6$, the Fe atoms have a $d^5$ state. However, in
Ba$_2$VFeO$_6$, we have an empty V $d$ shell. Injecting one electron
on V atoms changes its $d$ occupancy from $d^0$ to $d^1$, which does
not involve correlation effects. Injecting one electron on Fe atoms
changes its $d$ occupancy from $d^5$ to $d^6$, which increases the
number of electron pairs and each additional pair is associated with a
Hubbard $U$ energy. Therefore the V $d^0$ state is expected to lie
below the upper Hubbard band of Fe $d$ state, which means a smaller
gap for Ba$_2$VFeO$_6$ than for BiFeO$_3$.

Ref.~\cite{Chen2017} uses first-principles calculations to support the above
picture. In particular, the authors find that the polarization of
Ba$_2$VFeO$_6$ is comparable to that of BaTiO$_3$ and the gap of
Ba$_2$VFeO$_6$ is smaller than that of BaTiO$_3$ by about 1 eV. Since
the experimental optical gap of BaTiO$_3$ is 3.2 eV, it is predicted
that the optical gap of Ba$_2$VFeO$_6$ is around 2.2 eV, which is 0.5
eV smaller than the optical gap of BiFeO$_3$. This makes
Ba$_2$VFeO$_6$ a promising candidate among perovskite oxides for bulk
photovoltaic applications.

In addition to Ba$_2$VFeO$_3$, double perovskite Pb$_2$VFeO$_6$ has a
ferroelectric polarization comparable to PbTiO$_3$. Double perovskite
Sr$_2$VFeO$_6$, like SrTiO$_3$, is paraelectric but in the vicinity of
ferroelectric-paraelectric phase boundary. We note that in terms of
ferroelectric properties, $A_2$VFeO$_6$ has a simple one-to-one
correspondence to $A$TiO$_3$ ($A$=Ba, Pb, Sr).

\subsection{SrVO$_3$/SrMnO$_3$ and Sr$_2$VO$_4$/Sr$_2$MnO$_4$ superlattices }

\begin{figure}[t]
\includegraphics[angle=0,width=0.85\textwidth]{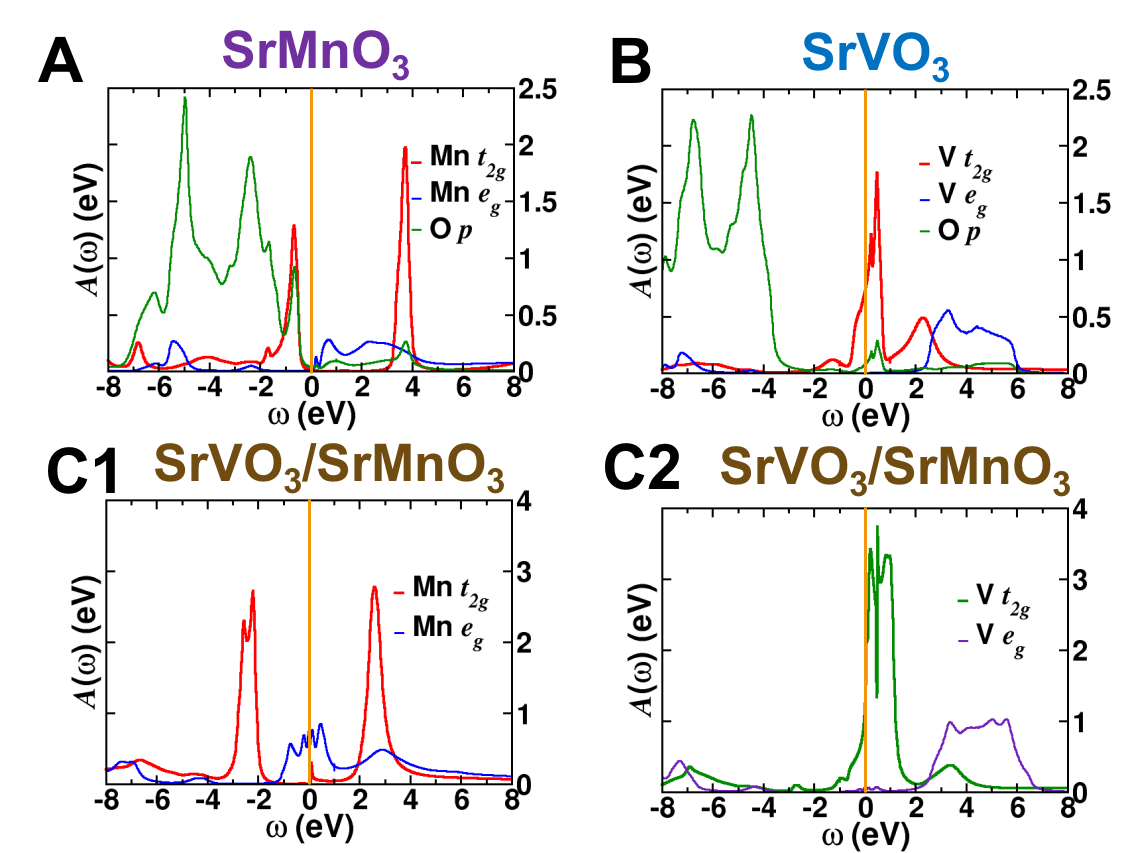}
\caption{\label{fig:VMn-dos} \textbf{A}) Spectral function of cubic
  SrMnO$_3$. The red, blue and green are Mn-$t_{2g}$, Mn-$e_g$ and
  O-$p$ projected density of states, respectively. \textbf{B})  Spectral function of cubic
  SrVO$_3$. The red, blue and green are V-$t_{2g}$, V-$e_g$ and O-$p$ 
  projected density of states, respectively. \textbf{C1}) Spectral
  function of SrVO$_3$/SrMnO$_3$ superlattices. The red and blue
  curves are Mn-$t_{2g}$ and Mn-$e_g$ projected density of states. \textbf{C2}) Spectral
  function of SrVO$_3$/SrMnO$_3$ superlattices. The green and purple
  curves are V-$t_{2g}$ and V-$e_g$ projected density of states. This
  figure is adapted from Ref.~\cite{Chen2014}.}
\end{figure}

In addition to the cases of nominally ``complete'' charge transfer (the formal
valence of cation ion changes by $\pm 1$) that are reviewed above, we
may also have partial charge transfer, if the electronegativity
difference between two similar transition metals is moderate. Partial
charge transfer generically leads to emergent metallic properties due
to the non-integer filling of bands.  Ref.~\cite{Chen2014} studies
SrVO$_3$/SrMnO$_3$ superlattices. Bulk SrVO$_3$ is a paramagnetic
metal with a nominal V $d^1$ occupancy, while bulk SrMnO$_3$ is an
antiferromagnetic insulator with a nominal Mn $d^3$ occupancy. In the
SrVO$_3$/SrMnO$_3$ superlattice, the partially occupied V $t_{2g}$
states have similar energy to the empty Mn $e_g$ states, which results
in an incomplete charge transfer, i.e. nominally the valence of V
changes from 4 to $(4+x)$ and the valence of Mn changes from 4 to
$(4-x)$ where $0< x <1$. Fig.~\ref{fig:VMn-dos} shows the
theoretically calculated spectral functions of bulk SrMnO$_3$,
SrVO$_3$ and (SrMnO$_3$)$_1$/(SrVO$_3$)$_1$ superlattice. All the
calculations are performed in paramagnetic
states. Fig.~\ref{fig:VMn-dos}\textbf{A} shows that a Mott gap is
opened in bulk SrMnO$_3$. Fig.~\ref{fig:VMn-dos}\textbf{B} shows that bulk SrVO$_3$ is
paramagnetic metallic with V $t_{2g}$ states at the Fermi surface. The
panels \textbf{C} of Fig.~\ref{fig:VMn-dos} show the spectral function
of (SrMnO$_3$)$_1$/(SrVO$_3$)$_1$ superlattice. In
Fig.~\ref{fig:VMn-dos}\textbf{C1}, the Mn $e_g$ states are partially
occupied, which leads to emergent metallic behavior on Mn atoms in the
superlattice. In Fig.~\ref{fig:VMn-dos}\textbf{C2}, the V $t_{2g}$
states are still partially occupied instead of empty, indicating that
nominally less than one complete electron is transferred from V to Mn,
in contrast to the complete charge transfer in
(LaTiO$_3$)$_1$/(LaNiO$_3$)$_1$ superlattice. Furthermore, since
SrMnO$_3$ is electron doped, the double exchange mechanism favors a
ferromagnetic ordering, just as in
La$_{1-x}$Sr$_x$MnO$_3$~\cite{Salamon2001}.  In the
(SrVO$_3$)$_1$/(SrMnO$_3$)$_1$ superlattice, ferromagnetism is
expected to emerge in the MnO$_2$ layer.

A closely related oxide heterostructure is Sr$_2$VO$_4$/Sr$_2$MnO$_4$
superlattice~\cite{Chen2014}. In this 214 Ruddlesden-Popper
superlattice, similar charge transfer phenomenon from V to Mn occurs
like the counterpart SrVO$_3$/SrMnO$_3$ superlattice. Synthesizing
transition metal oxides of a complicated Ruddlesden-Popper structure is now feasible in
experiment~\cite{Lee2013a}. Designing Ruddlesden-Popper superlattices is an
interesting direction for future research.

\subsection{manganite/cuprate interfaces}

\begin{figure}[t]
\includegraphics[angle=0,width=0.85\textwidth]{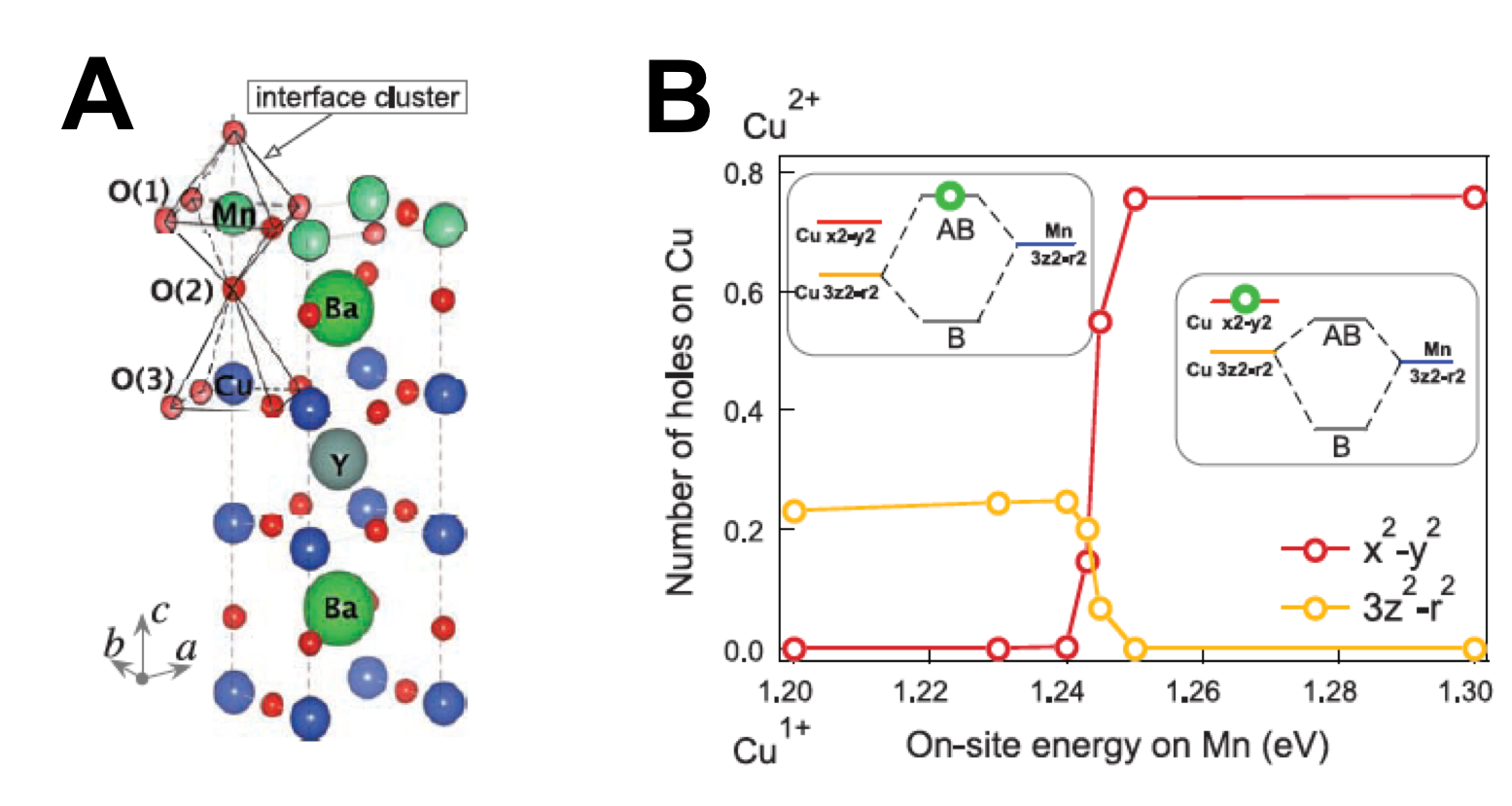}
\caption{\label{fig:Cu} \textbf{A}) Atomic positions near the
  La$_{2/3}$Ca$_{1/3}$MnO$_3$/YBa$_2$Cu$_3$O$_7$ (LCMO/YBCO) interface.  The
  MnCuO$_{10}$ cluster used for the exact-diagonalization calculations
  is highlighted. \textbf{B}) Occupancy of Cu $d$ orbitals at the
  LCMO/YBCO interface as a function of Mn hole on-site energy, as
  predicted by the exact-diagonalization calculations described in the
  text. The occupancy is given by the total number of holes, measured
  from the full-shell ($3d^{10}$) electron configuration. The corresponding
  formal Cu valence states are indicated for clarity.  The insets show
  the orbital level scheme at the interface, including extended
  bonding (B) and antibonding (AB) ``molecular orbitals'' formed by
  hybridized Cu and Mn $d_{3z^2-r^2}$ orbitals. The hole is indicated as the
  green circle. This figure is adapted from
  Ref.~\cite{Chakhalian2007}. }
\end{figure}

Partial charge transfer also occurs to the interface between the
ferromagnetic conducting manganite La$_{2/3}$Ca$_{1/3}$MnO$_3$ and the
superconducting cuprate YBa$_2$Cu$_3$O$_7$. The atomic structure of
the interface is shown in Fig.~\ref{fig:Cu}\textbf{A}. It is more complicated
than the $AM$O$_3$/$AM'$O$_3$ superlattices, but the underlying charge
transfer can be understood in a similar way to the previous examples
we have reviewed.

In Ref.~\cite{Chakhalian2007}, the authors observe a 0.2$e$ charge transfer from Mn to Cu
per ion pair across the interface between La$_{2/3}$Ca$_{1/3}$MnO$_3$
and YBa$_2$Cu$_3$O$_7$. The direction of charge transfer can be
deduced from our simple schematics of Fig.~\ref{fig:d-states}
that with respect to O-$p$ states, Mn-$d$ state lie above Cu-$d$
states. Fig.~\ref{fig:Cu}\textbf{B} shows a detailed analysis of
charge transfer at the interface. In theory, the on-site energy on Mn
can be considered as a tuning parameter to control the charge
transfer. In the right inset of Fig.~\ref{fig:Cu}\textbf{B}, the
energy of Mn $d_{3z^2-r^2}$ has the same energy as that of Cu
$d_{3z^2-r^2}$ and the hole resides on Cu $d_{x^2-y^2}$ orbital. In
the left inset of Fig.~\ref{fig:Cu}\textbf{A}, as the energy of Mn
$d_{3z^2-r^2}$ increases, a partial charge transfer occurs from Mn to
Cu. In addition, the antibonding state formed by Cu $d_{3z^2-r^2}$ 
and Mn $d_{3z^2-r^2}$ orbitals has higher energy than that of Cu
$d_{x^2-y^2}$ orbital. Therefore the hole on Cu atoms moves from
$d_{x^2-y^2}$ to $d_{3z^2-r^2}$ orbitals. In Ref.~\cite{Chakhalian2006}, at the same
interface, the authors also find significant re-arrangement of
magnetic domain structures accompanying charge transfer from Mn to Cu atoms.

\subsection{Antisite defects}

\begin{figure}[t]
\includegraphics[angle=0,width=0.85\textwidth]{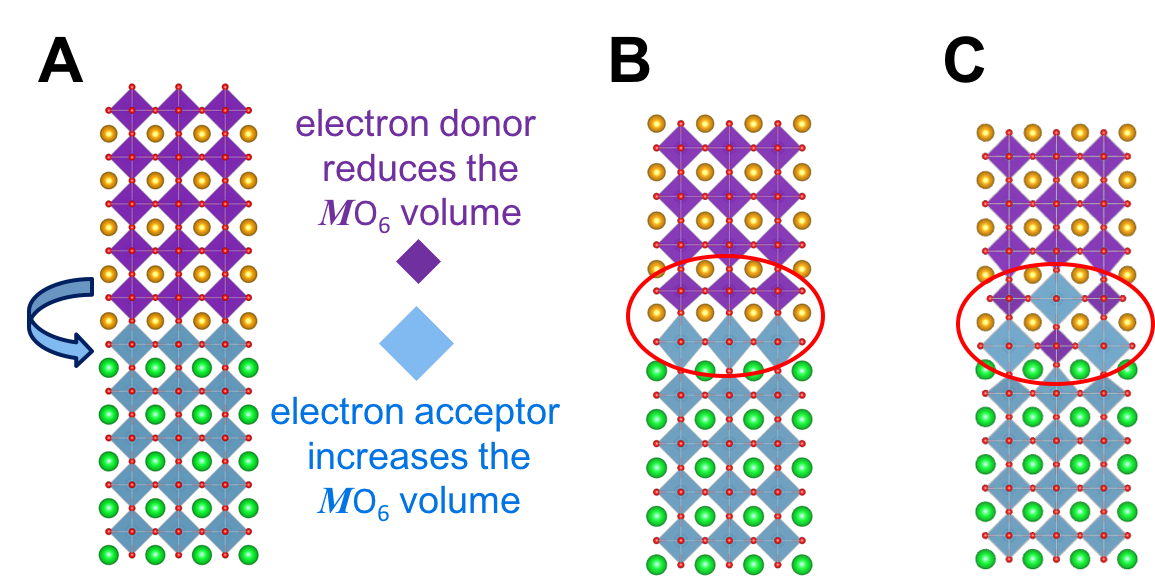}
\caption{\label{fig:antisite} \textbf{A}) Atomic structure of an 
interface between two semi-infinite perovskite oxides. The arrow
indicates a charge transfer. After the charge transfer, the electron
donor has a smaller $M$O$_6$ oxygen octahedron, while the electron
acceptor has a larger $M$O$_6$ oxygen octahedron. Here $M$ is a
transition metal. \textbf{B}) Atomic structure of an ideal interface
with substantial charge transfer across the interface. The $M$O$_6$
oxygen octahedron of electron donor is under tensile strain. The $M$O$_6$
oxygen octahedron of electron acceptor is under compressive
strain. \textbf{C}) Atomic structure of an interface with substantial
charge transfer across the interface and antisite defects.
The volume disproportionation of oxygen octahedron $M$O$_6$ between
electron donor and electron acceptor is naturally accommodated by
antisite defects.}
\end{figure}

In this sub-section, we discuss antisite defects at oxide interfaces,
which turn out to have close connections to charge
transfer~\cite{Chen2016, Neumann2012}.  Antisite defects in which atoms exchange
places across an interface may be important. Here we focus on one
particular type of antisite defect: at the interface between two
semi-infinite perovskite oxides, two $B$-site transition metal ions
interchange their positions. As Fig.~\ref{fig:antisite}\textbf{A}
shows, if substantial charge transfer occurs across the interface, the
$B$O$_6$ oxygen octahedron that donates the electron shrinks its
volume, while the $B'$O$_6$ oxygen octahedron that accepts the
electron expands its volume. If the interface remains atomically sharp
as in Fig.~\ref{fig:antisite}\textbf{B}, some $B$O$_6$ oxygen
octahedra (electron donors) are under tensile strain, while other
$B'$O$_6$ oxygen octahedra (electron acceptors) are under compressive
strain.  However, if antisite defects are induced at the interface
(see Fig.~\ref{fig:antisite}\textbf{C}), the volume disproportionation
will be naturally accommodated, which thus significantly reduces the
internal strain. This indicates that significant charge transfer
across oxide interfaces is a fundamental thermodynamic driving force
to induce antisite defects.

In Ref.~\cite{Chen2016}, the authors use first-principles methods to
survey 21 La$M$O$_3$/La$M'$O$_3$ interfaces ($M, M'$ = Ti, V, Cr, Mn,
Fe, Co, Ni) and 15 Sr$M$O$_3$/Sr$M'$O$_3$ interfaces ($M, M'$ = Ti, V,
Cr, Mn, Fe, Co). The authors find that about 50\% of the surveyed
interfaces have strong tendency for antisite defects and these
interfaces have a high degree of charge transfer between two
dissimilar transition metal ions. 

\begin{figure}[t]
\includegraphics[angle=0,width=0.85\textwidth]{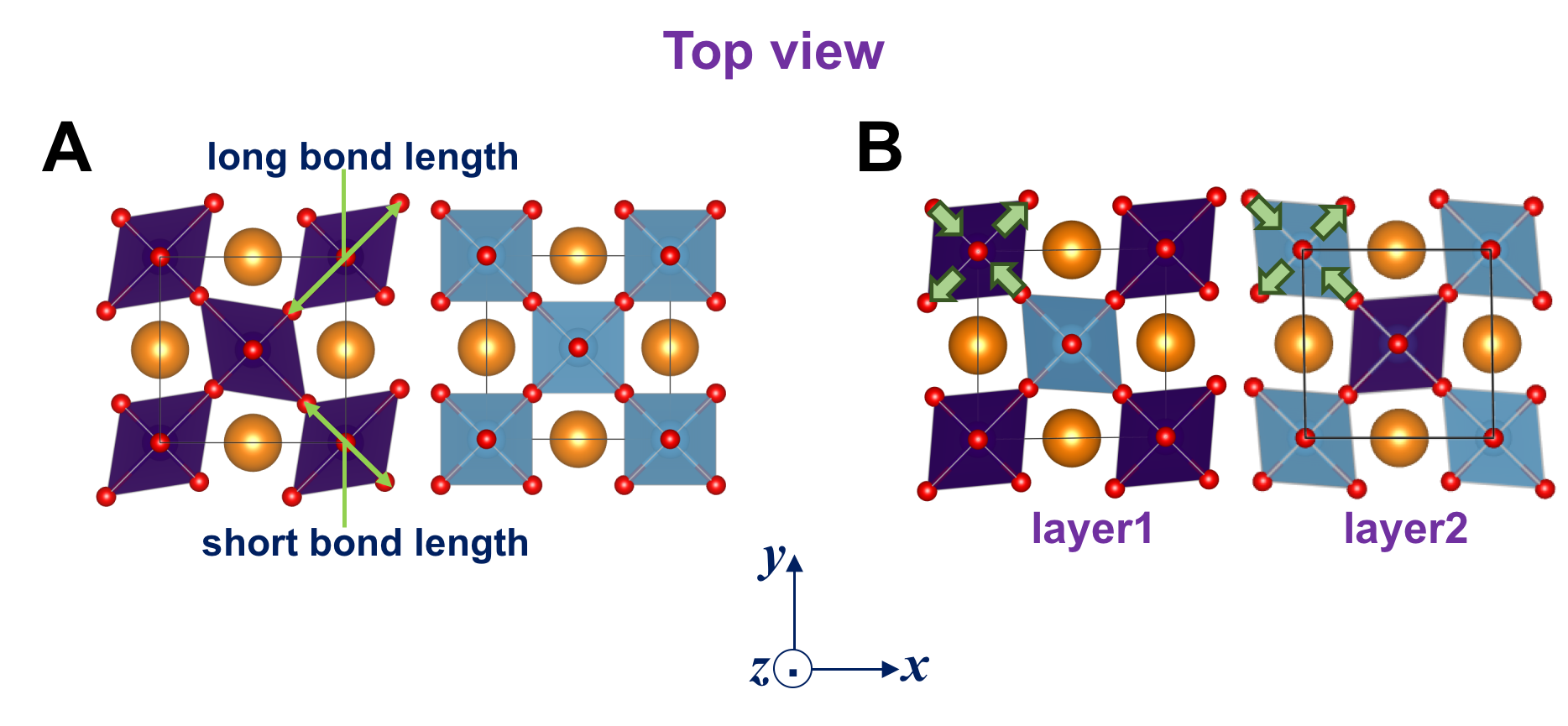}
\caption{\label{fig:Jahn-Teller} \textbf{A}) Top view of two
  vertically adjacent oxide layers at the interface without no antisite
  defects. The purple oxygen octahedron has strong Jahn-Teller
  distortions (one long metal-oxygen bond length and one short
  metal-oxygen bond length). The blue oxygen octahedron has no
  Jahn-Teller distortions (two metal-oxygen bond lengths are equal).
  \textbf{B}) Top view of two vertical adjacent oxide layers at the
  interface with one antisite defect. The purple oxygen octahedron has
  bond disproportionation (Jahn-Teller distortion) and the blue oxygen
  octahedron does not have bond disproportionation. Compatibility with
  the geometry imposes strains (green arrows) to reduce the bond
  disproportionation of the purple oxygen octahedron and to induce a
  disproportionation in the blue oxygen octahedron. Rotations and
  tilts of oxygen octahedra are suppressed for clarity. This figure is adapted from
  Ref.~\cite{Chen2016}.}
\end{figure}

Ref.~\cite{Chen2016} also shows that for interfaces with negligible
charge transfer, the presence of Jahn-Teller distortions can help
inhibit antisite defects. Fig.~\ref{fig:Jahn-Teller} shows the effects
of Jahn-Teller distortions at oxide interfaces. Panel \textbf{A})
shows the top view of two vertically adjacent oxide layers at the
interface without no antisite defects. The purple oxygen octahedron
has strong Jahn-Teller distortions (one long metal-oxygen bond length
and one short metal-oxygen bond length). The blue oxygen octahedron
has no Jahn-Teller distortions (two metal-oxygen bond lengths are
equal).  Panel \textbf{B}) shows the top view of two vertical adjacent
oxide layers at the interface with one antisite defect. The purple
oxygen octahedron has bond disproportionation (Jahn-Teller distortion)
and the blue oxygen octahedron does not have bond
disproportionation. Compatibility with the geometry imposes strains
(green arrows) to reduce the bond disproportionation of the purple
oxygen octahedron and to induce a bond disproportionation in the blue
oxygen octahedron. To reduce the elastic strain, the ideal interface
with no antisite defects is thermodynamically favored.

We summarize that antisite defects are strongly associated with
geometry constraints, which in turn are controlled by the charge
states of transition metal ions and are therefore closely connected to
charge transfer. Antisite defects are favored at oxide interfaces if
the defect allows the system to accomodate volume disproportionation
induced by charge transfer. On the other hand, if the ideal
(un-defected) interface can accomodate bond disproportionation due to
Jahn-Teller distortions (perhaps also due to charge transfer),
antisite defects are disfavored.

\section{Theoretical challenges}

In this section, we briefly review the theoretical methods that are
used to calculate oxide heterostructures and discuss the challenges
faced in order to better understand charge-transfer-driven phenomena
at oxide interfaces.

The key quantities to calculate are the band
alignments between occupied and unoccupied states in bulk
materials and between similar states on opposite sides of an interface. 
Therefore, the biggest challenge is to develop a method
with no fitting parameters that calculates electronic structure of
realistic materials (including complex heterostructures).

Currently, density functional theory (DFT)~\cite{Hohenberg1964, Kohn1965} with local density
approximation (LDA)~\cite{Perdew1981} and generalized gradient
approximation (GGA)~\cite{Perdew1996} 
is the workhorse to calculate the crystal structure of oxide
heterostructures. Because this method gives access to the energy as a
function of atomic positions, it can capture complicated distortions in oxides, including
oxygen octahedral rotations, ferroelectric
displacements and metal-oxygen bond disproportionation. However, DFT is a ground
state theory which (with the exact exchange correlation functional)
yields the correct ground state energy, its charge density
and crystal structure (after atomic relaxation). The DFT-calculated
electronic structure (band structure and density of state) that is
based on fictitious Kohn-Sham orbitals is in principle unphysical and
therefore band alignment need not be correct.
In practice, for weakly correlated materials such as band insulators,
the DFT-calculated electronic structure is qualitatively reasonable, but 
quantitatively it underestimates the size of band gaps by
30-50\%. However, for strongly correlated materials including
transition metal oxides, the DFT-calculated electronic structure can be qualitatively
incorrect (DFT predicts a metallic ground state for various Mott
insulators).  Since DFT can not accurately calculate energy
separation between metal $d$ and oxygen $p$ states in many strongly
correlated oxides, in oxide heterostructures DFT can also make
incorrect predictions on the band alignment of $d$ states between two
different transition metal atoms, which is the key variable to control
charge transfer phenomena.

In order to improve the electronic structure calculated by DFT,
various extensions and more sophisticated many-body theory methods
have been used in literature. One of the most widely used extension is
DFT plus Hubbard $U$ and Hund's $J$ corrections, commonly known as the
DFT+$U$ method~\cite{Liechtenstein1995, Dudarev1998}. In this method,
the correlation effects are treated in a static mean-field
approximation. The biggest advantage of this method is that its
computational scaling is almost the same as standard DFT calculations
and atomic relaxation can be performed within the method. However,
Hubbard $U$ and Hund's $J$ are element-dependent and are fixed
phenomenologically.  More importantly, DFT+$U$ method is a static mean
field approximation that can not describe many important dynamical
correlated phenomena, such as the Mott insulating state. DFT plus
dynamical mean field theory (DFT+DMFT) is another major extension of
DFT~\cite{Georges1996, Kotliar2006a}. In DFT+DMFT method, DFT
calculates the hopping matrix elements of the underlying lattice model
for realistic materials, while single-site DMFT calculates a
frequency-dependent self energy and the corresponding spectral
functions. DFT+DMFT method can describe many dynamical correlated
phenomena, such as Mott state and correlation-driven band
reduction. More importantly, if Hubbard $U$ and Hund's $J$ parameters
are correct, the band alignment based on the DMFT-calculated spectral
functions is more accurate and reliable than that based on the
DFT-calculated density of states. However, like DFT+$U$, DFT+DMFT
method itself does not calculate the element-dependent $U$ and
$J$. Furthermore, the calculation of forces on atoms in
complex solids within the DMFT method is still in
infancy~\cite{Chakhalian2014, Leonov2014, Haule2016}. Therefore unlike DFT+$U$ method, atomic
relaxation is not feasible in DFT+DMFT method at this stage. Another
important issue in both DFT+$U$ and DFT+DMFT is the double counting
problem. In both methods, the separation of DFT and extension raises
the possibility that some interactions will be included in both parts
and will therefore be counted twice, necessitating the subtraction of
an additional “double counting” term.  The physical properties
calculated from DFT+$U$ or DFT+DMFT sensitively depend on double
counting, but unfortunately the exact form of double counting is
unknown. In literature a widely used empirical double counting form is
called fully localized limit (FLL)~\cite{Czyzyk1994}. However, recent
work~\cite{Park2014} shows that the FLL double counting may lead to an inaccurate
energy separation between metal $d$ and oxygen $p$ states in rare
earth nickelates. However, how to improve the FLL counting in
DFT+$U$ and DFT+DMFT calculations is one of the biggest theoretical challenges of the methods.

Another two methods--hybrid functional~\cite{Kummel2008} and
GW~\cite{Hedin1965}--have also been used in
literature to calculate the electronic structure of complex
oxides as an improvement over DFT. The advantage of both methods is
that they do not involve material-dependent parameters but
both methods are very computationally intensive. Therefore the calculations
using both methods are constrained to small systems and atomic relaxation
is not practically feasible for complex heterostructures.

We note that in many strongly correlated materials, electronic structure
and atomic structure are closely related. For example, VO$_2$
undergoes a coupled metal-insulator rutile-monoclinic
transition~\cite{Morin1959, Barker1966}. While it is still an on-going research topic whether the
transition is primarily driven by electronic transition or structural
transition, it is a classical example for strongly correlated
materials that different atomic structures correspond to distinct
electronic structures. For oxide heterostructures, atoms close to the
interface generically move away from their positions in bulk
constituents and charge transfer phenomenon is strongly coupled to the
new atomic positions because they can significantly change the energy
separation between metal $d$ states and oxygen $p$ states as well as
hopping matrix elements and band widths~\cite{Chen2016}. Currently DFT and DFT+$U$
methods can efficiently calculate forces on atoms in solids and
therefore can perform atomic relaxation and obtain optimal atomic
positions for complex heterostructures. However, strong correlation
effects are either neglected in DFT or treated in a static mean field
approximation in DFT+$U$. On the other hand, DMFT/hybrid functional/GW
improve the calculations of electronic structure to different extent,
but atomic relaxation is very difficult, if not possible, using these
sophisticated methods. The compromising approach of using DFT/DFT+$U$
to obtain the optimal atomic structure or simply using experimentally
determined atomic structure, and then using DMFT/hybrid functional/GW
methods to calculate electronic structure is currently preferred. A unified theory which can
calculate both electronic and atomic structures for strongly correlated
materials on the same footing is highly desirable but very
challenging. We finally note that while the calculation of
many-body band offsets is a key theoretical challenge, measurements of
charge transfer and band offsets in experiment can provide a key test of theories.

\section{Summary and Perspectives}

We reviewed three major mechanisms for charge transfer in oxide
heterostructures. In our classification, charge transfer can occur
across oxide interfaces in order to compensate for 1) polarity
difference, 2) occupancy difference and 3) electronegativity difference
between two different transition metal oxides. We summarized
representative examples for the first two mechanisms and present a
more detailed review of important examples for the third
mechanism. Table~\ref{tab:summary} provides a quick summary.
We also reviewed the theoretical methods used to study charge
transfer phenomena in oxide heterostructures and discuss the
challenges we face in theory. 

Oxide heterostructures have shown a plethora of properties which are
not exhibited in their bulk constituents. Charge transfer is a very
general and robust phenomenon that occurs to oxide interfaces.  In the
review, we highlight oxide interfaces in which charge transfer occurs
to $3d$ transition metal ions. However, recent experimental progress
makes it feasible to synthesize oxide heterostructures that contain
$4d$ and $5d$ transition metal ions~\cite{Matsuno2015}. Charge
transfer between $3d$-to-$4d$ or $3d$-to-$5d$ transition metal ions is
a very interesting direction for future research, since spin-orbit
interaction is stronger in $4d$ and $5d$ transition metal ions and the
interplay between correlation effects and spin-orbit interaction will
play a crucial role in charge transfer phenomena. While our review
mainly focuses on (001) interfaces, the emergent phenomena identified in
oxide heterostructures may also be present in
mixed bulk materials, such as double perovskite oxides (e.g. see
Section V\textbf{D}). We hope our review
can stimulate further theoretical and experimental work to search for
novel strongly correlated phenomena in oxide heterostructures.

\begin{acknowledgments}
We are grateful to useful discussion with M. Bibes, H. T. Dang,
S. Dong, A. Georges, M. Gibert, C. Marianetti, H. Park, K. Shen,
D. Schlom, Z. Zhong and in particular S. Ismail-Beigi.
H. Chen is supported by National Science Foundation under Grant
No. DMR-1120296. A. J. Millis is supported by the Department of Energy
under Grant No. DOE-ER-046169. Computational facilities are provided
via Extreme Science and Engineering Discovery Environment (XSEDE),
which is supported by National Science Foundation through Grant
No. TG-PHY130003 and also via the National Energy Research Scientific
Computing Center (NERSC), a DOE Office of Science User Facility
supported by the Office of Science of the U.S.  Department of Energy.
\end{acknowledgments}

\clearpage
\newpage

\begin{table}[h!]
\caption{\label{tab:summary} Table of different charge transfer
  mechanisms in oxide heterostructures. For each mechanism,
  representative examples with the corresponding emergent phenomena
  are provided.} 
\begin{center}
%\begin{tabularx}{\textwidth}{c *{3}{|Y} c}
\begin{tabularx}{\textwidth}{p{0.14\textwidth}|p{0.18\textwidth}|p{0.66\textwidth}}
\hline\hline
Mechanisms  &   Examples  &  Emergent phenomena\\
\hline
Polarity difference  &  LaAlO$_3$/SrTiO$_3$  & The interface is metallic, magnetic and superconducting, although the constituents are insulators in bulk.\\
\hline
Occupancy difference & LaTiO$_3$/SrTiO$_3$  &  The
                                                                    interface
                                                                    is
                                                                    metallic, although LaTiO$_3$ is a Mott
                                              insulator and SrTiO$_3$
                                   is a band insulator
                                                                    .\\
\hline
\multirow{ 6}{0.14\textwidth}{Electro-negativity difference} & LaTiO$_3$/LaNiO$_3$ & Ni at the
                                                      interface is in a
                                                         $d^8$ Mott
                                                         insulating
                                                         state, although Ni in LaNiO$_3$ is
                                                      in a $d^7$
                                                      metallic state.
                                                      \\
\cline{2-3}
                                          & 
                           LaTiO$_3$/LaNiO$_3$ /LaAlO$_3$  &
                                                             Ni at
                                                             the
                                                             interface
                                                             has a
  huge orbital polarization, although Ni in LaNiO$_3$
                                              has a negligible orbital
                                              polarization.\\
\cline{2-3}
                                   &   LaMnO$_3$/LaNiO$_3$ &
                                                             The interface can be either
                                                insulating or metallic
  depending on the thickness of LaMnO$_3$ and LaNiO$_3$, although LaMnO$_3$
                                                             is an insulator
                                                             and 
                                                             LaNiO$_3$
                                                             is 
                                                      a metal.\\
\cline{2-3}
             &  Ba$_2$VFeO$_6$  & Ba$_2$VFeO$_6$
                                                               is
                                                               ferroelectric although both BaVO$_3$ and BaFeO$_3$ have cubic
                                  structures (not ferroelectric).\\
\cline{2-3}
             &  SrVO$_3$/SrMnO$_3$  & Mn at the interface is
                                                                doped
                                                                and becomes
                                                                metallic although SrMnO$_3$ is an insulator.\\
\cline{2-3}
    &   manganite /cuprate &  At the
                             interface the Cu has a
                                                               multi-orbital
                                                               Fermi
                                                               surface, although it is single-band in bulk;
                                                               and the Mn forms
                                                               different
                                                               magnetic
                                                               domain structures from bulk.\\
\hline\hline
\end{tabularx}
\end{center}
\end{table}

\clearpage
\newpage

%\bibliography{charge-transfer-review-v9}

\end{document}